\RequirePackage{fix-cm}

\documentclass[smallextended]{svjour3}       % onecolumn (second format)

\usepackage{amsmath,amssymb}
\usepackage{graphicx}
\usepackage{color}
\usepackage{mathrsfs}
\usepackage{bm}
\usepackage{multirow}
\usepackage{booktabs}
\usepackage{comment}
\usepackage{authblk}

\newcommand{\Part}[3]{ \frac{ \partial^{#3} #1 }{ \partial #2^{#3} } }%partial derivative
\newcommand{\V}[1]{\bm{#1} } %vector command
\newcommand{\Tr}[1]{ \mathop{\rm Tr}_{ #1 } }
\newcommand{\Ave}[1]{\left\langle {#1} \right\rangle} %thermal average 
\newcommand{\mN}{\mathbb{N}}
\newcommand{\lb}{\left(}
\newcommand{\rb}{\right)}
\newcommand{\lbb}{\left\{}
\newcommand{\rbb}{\right\}}
\newcommand{\lsb}{ \left[ }
\newcommand{\rsb}{ \right] }

\newcommand{\Req}[1]{eq.\ (\ref{eq:#1})}
\newcommand{\BReq}[1]{Eq.\ (\ref{eq:#1})}
\newcommand{\NReq}[1]{(\ref{eq:#1})}
\newcommand{\Reqs}[2]{eqs.\ (\ref{eq:#1},\ref{eq:#2})}

\newcommand{\Reqss}[2]{eqs.\ (\ref{eq:#1}-\ref{eq:#2})}

\newcommand{\Rfig}[1]{Fig.\ \ref{fig:#1}}

\newcommand{\NRfig}[1]{\ref{fig:#1}}
\newcommand{\Lfig}[1]{\label{fig:#1}}
\newcommand{\Leq}[1]{\label{eq:#1}}
\newcommand{\Rsec}[1]{sec.\ \ref{sec:#1}}
\newcommand{\BRsec}[1]{Sec.\ \ref{sec:#1}}

\newcommand{\Lsec}[1]{\label{sec:#1}}
\newcommand{\be}{\begin{eqnarray}}
\newcommand{\ee}{\end{eqnarray}}
\newcommand{\subbe}{\begin{subequations}}
\newcommand{\subee}{\end{subequations}}
\newcommand{\ba}{\begin{array}}
\newcommand{\ea}{\end{array}}
\newcommand{\no}{\nonumber}

\newcommand{\bc}{\begin{comment}}
\newcommand{\ec}{\end{comment}}

\title{Learning probabilities  from random observables in high dimensions: \\ the maximum entropy distribution and others}

\titlerunning{Learning probabilities: maximum entropy and others}        % if too long for running head

\author{Tomoyuki Obuchi   \and Simona Cocco \and  R\'emi Monasson}

\institute{
T. Obuchi \at Department of Computational Intelligence and Systems Science, Tokyo Institute of Technology, Yokohama, Japan \\
\email{obuchi@sp.dis.titech.ac.jp}           %  \\
 \and
S. Cocco \at Laboratoire de Physique Statistique de l'Ecole Normale Sup\'erieure, affili\'e au CNRS \& \`a l'Universit\'e Pierre et Marie Curie, Paris, France
\and
R. Monasson \at Laboratoire de Physique Th\'eorique de l'Ecole Normale Sup\'erieure, affili\'e au CNRS \& \`a l'Universit\'e Pierre et Marie Curie, Paris, France
}

\begin{document}

\maketitle

\begin{abstract}
We consider the problem of learning a target probability distribution over a set of $N$ binary variables from the knowledge of the expectation values (with this target distribution) of $M$  observables, drawn uniformly at random. The space of all probability distributions compatible with these $M$ expectation values within some fixed accuracy, called version space, is studied. We introduce a biased measure over the version space, which gives a boost increasing exponentially with the entropy of the distributions and with an arbitrary inverse `temperature' $\Gamma$. The choice of $\Gamma$ allows us to interpolate smoothly between the unbiased measure over all distributions in the version space ($\Gamma=0$) and the pointwise measure concentrated at the maximum entropy distribution ($\Gamma \to \infty$). Using the replica method we compute the volume of the version space and other quantities of interest, such as the distance $R$ between the target distribution and the center-of-mass distribution over the version space, as functions of $\alpha=(\log M)/N$ and $\Gamma$ for large $N$. Phase transitions at critical values of $\alpha$ are found, corresponding to qualitative improvements in the learning of the target distribution and to the decrease of the distance $R$. However, for fixed $\alpha$, the distance $R$ does not vary with $\Gamma$, which means that the maximum entropy distribution is not closer to the target distribution than any other distribution compatible with the observable values. Our results are confirmed by Monte Carlo sampling of the version space for small system sizes ($N\le 10$).   
\end{abstract}

%\newpage
%\tableofcontents

%%%%%%%%%%%%%%%%%%%%%%%%%%%%%%%%%%%%%%%%%%%%%%%%%%%%%%%%%%%%%%%%%%%%%%%%%%%%%%%%%%
\section{Introduction}

Multi-components and strongly interacting systems, in physics and beyond, may show complex behaviours eluding simple quantitative modeling. A common strategy to describe such systems is to define probability distributions over the space of their configurations. The task is at first sight daunting. The number of unknown probabilities scales as the dimension of the configuration space, and is enormous, generally exponentially large in the number of the degrees of freedom defining the system. The selection of one probability distribution among the multitude of possibilities may be done in a Bayesian way. A class of parametrized model distributions is considered, and a good choice of the parameters is sought, {\em e.g.} which maximizes the likelihood of the observed data. Statisticians are generally interested in understanding how learning proceeds, that is, in the speed of convergence to the target distribution (assumed to be in the parametrized class, and to be consistent with one value of the parameters) as more and more data are made available. Yet, though the number of parameters defining the class of distributions may be very large and the inference problem may be computationally very hard, the classes of parametrized distributions are generally extremely small compared to the plethora of possible distributions over the configuration space. While some classes of distributions may appear more adequate to represent the data or more amenable to computations, those criteria are largely arbitrary. 

The Maximum Entropy (ME) principle is an alternative approach, rejecting the notion of arbitrariness. The ME principle may be informally stated as follows: among all possible distributions compatible with what is known of the data, choose the one with ME. It was proposed as an alternative foundation of statistical mechanics \cite{Jaynes:57,Balian:94}; as an illustration, the Boltzmann distribution in the canonical ensemble may be found back as the distribution with ME under the sole knowledge of the average value of the energy. The ME principle is supported by information theory \cite{cover}, which argues that any other distribution would be too constrained, and, in fact, reflect additional properties of the data \cite{Jaynes:82}. An alternative and somewhat colourful formulation of this argument was given in \cite{Jaynes:86}, where the ME distribution was shown to emerge as the most frequent distribution which an uninformed operator (monkeys in \cite{Jaynes:86})  would find, upon repeated and unbiased trials compatible with the observations on the data. In other words, given an unbiased prior over the set of all possible distributions over the configuration space, the ME distribution is the most likely one compatible with our knowledge of the data. Furthermore the ME distribution enjoys important and valuable properties, {\em e.g.} a weak sensitivity to measurement errors \cite{Tikochinsky:84}. To the practitioner, the most compelling argument in support of the ME distribution may actually come from the successful applications to experimental data, see for instance \cite{Bialek:book} for a clear  presentation regarding biological data.

The purpose of the present work is to compare the performances of the ME distribution with the ones of the other distributions compatible with the data. While the literature on the ME principle has a rich history, we are not aware of works attempting to carry out this comparison in a quantitative and rather general setting. We consider a set of $N$ variables, taking binary values $\pm 1$ (the generalization to a larger number of states would be straightforward, as long as it remains finite). Any distribution over the space of configurations is entirely characterized by $2^N$ probabilities, which are non-negative real numbers summing to unity. Each distribution may therefore be seen as a point in the $2^N$--dimensional simplex. We pick up one point in this simplex, hereafter referred to as the target distribution. Next we consider a set of $M$ observables, each of which may be any polynomial function of the variables, and compute their average values over the target distribution. The distributions in the simplex compatible with those data, {\em i.e.} such that the observables have the same average within some prescribed accuracy,  are called admissible. The set of admissible distributions, called version space, contains the target distribution, the ME distributions, and many others, such as the center-of-mass distribution, which is the flat average over all admissible distributions. Our objective is to compute the main geometrical features of the version space, {\em e.g.} its volume, the distances from the target to the ME or to the center-of-mass distributions, the average inter-distribution distance, ....

To give a precise meaning to those quantities in a mathematically tractable framework we will assume that observables are drawn from a simple statistical ensemble. More precisely  the  values taken by the observables are assumed to be random and uncorrelated across the $2^N$ configurations. This assumption is not meant to be realistic. In most real applications, indeed, observables reflect the low-order statistics of the configurations, {\em e.g.} the value of the first variable, the product of the fifth and seventh variables, .... Such observables vary very smoothly over the configuration space, as they depend only a small (compared to $N$) number of the configuration variables, and may be adequate to provide information about smooth distributions. Our hypothesis can be considered as worst-case-like in the following sense. The inference of the target distribution from the average values of the observables may be recast as the problem of reconstructing a $2^N$--dimensional non-negative vector from the knowledge of its scalar products with $M$ vectors (corresponding to the observables). When those vectors are randomly chosen the scalar products are typically very small, of the order of $2^{-N/2}$, and are weakly informative about the direction of the target vector. In this pessimistic setting we expect that the number of scalar products (observables) necessary to reconstruct the target vector with accuracy will be of the order of the number of relevant components of the target vector, that is, of the order of the exponential of the entropy of the target distribution. While this statement is correct we will see that some important features of the target distribution are correctly inferred even with a much smaller number $M$ of observables. In particular, the probabilities of configurations with large values of the target distribution are learned with a limited number of data, a phenomenon connected to the onset of phase transitions in the learning process.

An important aspect of our framework is that it allows us to bias the measure over the version space, in order to boost the distributions with large entropies. The magnitude of this entropic bias may be continuously tuned from zero (uniform measure over the version space) to infinity, which amounts to selecting the ME distribution alone. We study how the distance between the target distribution and center-of-mass distribution varies with the bias (for a fixed number $M$ of observables). Our main result is that this distance does not depend on the bias, showing that the ME entropy is not better than any other distribution randomly picked up in the version space. While this result is valid for any target distribution in the case of random observables we do not expect it to apply to the more realistic case where both the observables and  the target distribution are smooth. 

All our results are derived within the replica symmetric framework when $N$ and $M$ go to infinity at fixed ratio $\alpha=(\log M)/N$, and are therefore non rigorous. We have, however, checked the local stability of our replica-symmetric solution against replica-symmetry-breaking fluctuations, the so-called replicon modes. Our results are therefore self-consistent and we expect them to be correct for large $N,M$. In addition we have designed a Monte Carlo algorithm to sample the version space, and applied it to small system sizes ($N\le 10$). Remarkably, simulations show only weak finite-size effects compared to our large--$N$ calculations; a good qualitative (and sometimes even quantitative) agreement with our analytical predictions is found.

The paper is organized to be accessible to the reader not interested in the details of our calculation. In \Rsec{sec89}, we present the necessary definitions and notations. An overview of our results, free of technicalities, is given in \Rsec{overview}. All technical details and calculations are reported in \Rsec{analysis} and in the Appendix.  We present the sampling algorithm and the results of our numerical simulations in \Rsec{numerics}. Conclusions can be found in \Rsec{concl}.

%%%%%%%%%%%%%%%%%%%%%%%%%%%%%%%%%%%%%%%%%%%%%%%%%%%%%%%%%%%%%%%%%%%%%%%%%%%%%%%%%%
\section{Definitions and notations}\Lsec{sec89}

%%%%%%%%%%%%%%%%%%%%%%%%%%%%%%%%%%%%%%%%%
\subsection{Target distribution }

Let us consider a system consisting of $N$ Ising spins, with configurations $\V{s}=\{s_i=\pm 1\}_{i=1}^{N}$. The probability distributions of the system configurations, hereafter called target distribution, is denoted by $\hat{p}_{\V{s}}$. 
%Our purpose is to infer $\{\hat{p}_{\V{s}}\}_{\V{s}}$ based on the knowledge of the observable values. 
We consider large-size systems, $N\gg 1$, and write
\be
\hat{p}_{\V{s}}\doteq e^{-N\omega_{\V{s}}},
\ee
where the rate $\omega_{\V{s}}>0$ of the configuration probability $\hat{p}_{\V{s}}$ is introduced, and the symbol $\doteq$ stands for equality in the leading exponential--in--$N$ term. It is convenient to introduce the entropy $\sigma$ of the rates $\omega_{\V{s}}$,
\be
\sigma\lb \omega \rb=\lim _{\epsilon\to 0^+} \lim_{N\to\infty} \frac{1}{N} \log \left[ \text{nb. of configurations $\V{s}$ such that } e^{-N\omega } \le\hat{p}_{\V{s}} < e^{-N ( \omega -\epsilon) }
\right] \ ,
\ee
The entropy curve $\sigma(\omega)$ has some remarkable features  in standard physical systems. First, it is convex and bounded from above and below. Secondly, the maximum of the curve is $\sigma_{0}=\log 2$; the corresponding value of $\omega$ is denoted by $\omega_{0}$. Thirdly, the curve lies below the $\sigma=\omega$ line, and is tangent to this line at $\omega_{1}$($\le\omega_0$). The value $\sigma_{1}=\sigma(\omega_1)$ is the entropy per spin of the target distribution
\be
\sigma_1= \lim_{N\to\infty}-\frac{1}{N}\sum_{\V{s}}\hat{p}_{\V{s}}\log \hat{p}_{\V{s}}.
\ee
More generally, given a real number $k$, we define $\omega_{k}$, $\sigma_{k}$, and $\ell_k$ through 
\be\Leq{lk}
\left. \frac{d\sigma}{d\omega}\right|_{\omega=\omega_k}=k\ , \quad 
\sigma_{k}\equiv \sigma(\omega_{k}) \ ,  \quad \ell_{k} \equiv k\, \omega_{k}-\sigma_k \ .
\ee
Note that the range of values of $k$ such that $\omega_{k}$ is well-defined is generally bounded. These quantities characterize the dominant contribution to the $k$th moment  of $\hat{p}$:
\be
\sum_{\V{s}}(\hat{p}_{\V{s}})^{k}
=\int d\omega\, e^{N\lb \sigma(\omega)-k\, \omega\rb} \doteq e^{N\lb \sigma_k-k\, \omega_k\rb} = e^{-N\ell_{k}} \ .
\ee
As an illustration of the properties above we show in \Rfig{entropy-curve} the entropy curve of the independent spin model (ISM) defined as
\be
\hat{p}^{\rm ISM}_{\V{s}}=\frac{1  }{(2\cosh H)^N}\; \exp\left(H\, \sum_{i=1}^Ns_{i}  \right)\ ,
%\frac{e^{H(N-2N_{s}^{-})}  }{(2\cosh H)^N}=\hat{p}^{\rm ISM}_{N^{-}_{\V{s} } },
\Leq{ISM}
\ee
%where $N^{-}_{ \V{s} }$ is the number of down spins. 
for $H=0.5$. The value of $H$ will not change throughout this paper.
\begin{figure}[htbp]
\begin{center}
\includegraphics[width=0.60\columnwidth]{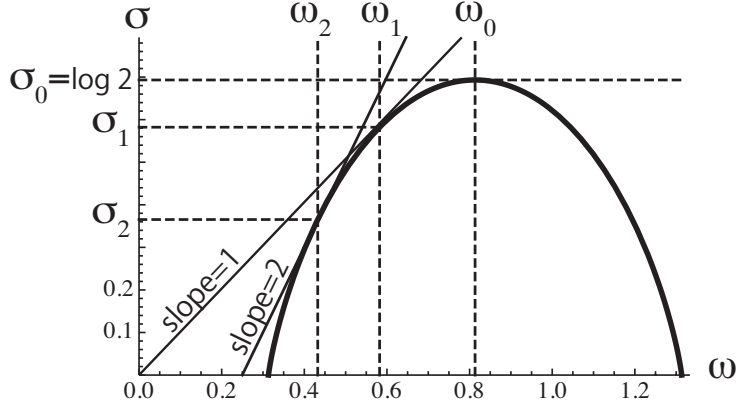}
\caption{Entropy $\sigma$ as a function of $\omega$ for the independent spin model in \Req{ISM} with $H=0.5$. The entropy curve is obtained through a parametric representation $\omega(m)=\log( 2\cosh H) -Hm$, $\sigma(m)=\log 2 -\frac 12 (1+m)\log(1+m)-\frac 12 (1-m)\log(1-m)$, where $m$ is the magnetization per spin, ranging from $-1$ to $+1$. It is easy to show that $\omega_k=\omega(\tanh (kH))$.
}
\Lfig{entropy-curve}
\end{center}
\end{figure}

%%%%%%%%%%%%%%%%%%%%%%%%%%%%%%%%%%%%%%%%%
\subsection{Observables, version space, and maximum entropy distribution}

Let $\V{v}$ be an observable, taking value $v_{\V{s}}$ when the system configuration is $\V{s}$. Observables and probability distributions can be seen as vectors in a $2^N$--dimensional space, with components labelled by the configurations $\V{s}$. We assume that measurements give us access to the average value of the observable over the target distribution,
\be\Leq{Obs}
\sum_{ \V{s} } v_{ \V{s} } \,\hat{p}_{\V{s}}\equiv \V{v}\cdot \hat{\V{p}},
\ee
which may be simply written as the scalar product of the vectors attached to the observable and to the target distribution. Suppose we have made $M$ measurements corresponding to $M$ observables $\V{v}^{\mu}\, (\mu=1,\cdots,M)$. An `admissible' probability vector, $\V{p}$, compatible with all the measurements is such that 
\be
 \lb \V{p}-\hat{\V{p}} \rb \cdot \V{v}^{\mu}=0 \ , \quad \forall\ \mu=1,\cdots,M.
\Leq{constraint}
\ee 
We hereafter use the term `constraints' to refer to \Req{constraint} or to the attached observables $\V{v}^{\mu}$. The set of vectors satisfying \Req{constraint}, together with the normalization and the non-negativity conditions
\be\Leq{normcond}
\sum_{\V{s}}p_{\V{s}}=1\quad \text{and}\quad p_{\V{s}}\geq 0\ , \forall{\V{s}} \ ,
\ee
defines  the version space.

Each distribution $\V{p}$ in the version space is characterized by its Shannon entropy,  
\be \Leq{sha}
S(\V{p})=-\sum_{\V{s}}p_{\V{s}}\log p_{\V{s}} \ ,
\ee
which may be interpreted as an estimate of the logarithm of the effective number of configurations under the distribution. The maximum entropy (ME) distribution, $\V{p}_{ME}$, is the distribution maximizing \Req{sha} in the version space, {\em i.e.} under the constraints listed in \Req{constraint} and in \Req{normcond}. Using Lagrange multipliers $\eta^\mu$ to enforce those constraints the ME distribution may be formally written as
\be
p^{\rm ME}_{\V{s}} = C\; \exp\lb \sum_{\mu}\eta^{\mu}\, v^{\mu}_{\V{s}} \rb,
\Leq{dist-ME1}
\ee
where $C$ is a normalization constant. As an illustration, if we consider the set of the $N$ single-spin observables, $v_{\V{s}}= s_i $, $1\le i\le N$, and of the $N(N-1)/2$ two-spin observables $v_{\V{s}}= s_i \, s_j$, $1\le i < j\le N$, we recover the well-known result that the Ising model is the ME model given the average values of those observables. The Lagrange multipliers $\eta$ coincide with, respectively, the $N$ fields and the $N(N-1)/2$ pairwise couplings acting on the spins.

%%%%%%%%%%%%%%%%%%%%%%%%%%%%%%%%%%%%%%%%%
\subsection{Measures over the space of distributions}

In realistic situations involving certain measurement noise, it may be beneficial not to perfectly reproduce the given average value by avoiding overfitting. To take into account such a flexibility, we introduce a Gaussian-like measure with variance $E$ over the vector space of $\V{p}$:
\be\Leq{measure}
\rho\lb \V{p}|\{ \V{v}^{\mu}\}_{\mu=1}^{M},\hat{\V{p}} \rb=\frac{1}{V}
 \prod_{\V{s}}\theta(p_{\V{s}}) \,
  \delta\lb \sum_{ \V{s}}p_{\V{s}} -1  \rb 
\exp \lbb 
-\frac{1}{2E}\sum_{\mu=1}^{M} \lb \V{v}^{\mu}\cdot \lb \V{p}-\hat{\V{p}} \rb \rb^{2} 
\rbb,
\ee
where $\theta(x)$ is the step function, equal to $1$ if $x\geq 0$ and to $0$ if $x<0$, and $\delta(x)$ is the Dirac delta function. We will call $E$ tolerance hereafter. The denominator $V$ is defined to make sure that the measure $\rho$ is normalized: 
\be \Leq{volume1}
V \lb E, \{ \V{v}^{\mu}\}_{\mu=1}^{M},\hat{\V{p}} \rb
=\int_{0}^{\infty}\prod_{\V{s}}dp_{\V{s}}\, \delta\lb \sum_{ \V{s}}p_{\V{s}} -1  \rb 
\exp \lbb 
-\frac{1}{2E}\sum_{\mu=1}^{M} \lb \V{v}^{\mu}\cdot \lb \V{p}-\hat{\V{p}} \rb \rb^{2} 
\rbb.
\ee
This normalization factor measures the volume of `admissible'   probability vectors given the constraints \Req{constraint}.  In this probabilistic setting we will loosely use the term `version space' to refer to the set of distributions $\V{p}$ associated to `large'  measure values $\rho(\V{p})$. Note that, while $\rho(\V{p})$ defines the joint-measure of the $2^N$--configuration probabilities, we will also consider below the marginal measure for a single-configuration probability, 
\be \Leq{marginal}
\rho_{\V{s}}(p_{\V{s}}|\{ \V{v}^{\mu}\}_{\mu=1}^{M},\hat{\V{p}})=\int_{0}^{\infty}\prod_{\V{t}(\neq \V{s})}dp_{\V{t}} \, \rho(\V{p}|\{ \V{v}^{\mu}\}_{\mu=1}^{M},\hat{\V{p}}) \ .
\ee

It is possible to consider other measures of the space of $\V{p}$ for the purpose of studying the performances of the ME distribution.  To favor the probability vectors $\V{p}$ with large Shannon entropies $S(\V{p})$, see \Req{sha}, it is natural to introduce the new measure
\be
\rho\lb \V{p}|\Gamma, \{ \V{v}^{\mu}\}_{\mu=1}^{M},\hat{\V{p}} \rb \propto \rho\lb \V{p}|\{ \V{v}^{\mu}\}_{\mu=1}^{M},\hat{\V{p}} \rb\; e^{\Gamma\; S(\V{p})}\ ,
\ee
where $\Gamma$ is the strength of this entropic bias. The normalization of this new measure $\rho$ implicitely defines the following expression for the volume in the presence of an entropic bias, 
\be
V \lb\Gamma,E, \{ \V{v}^{\mu}\}_{\mu=1}^{M},\hat{\V{p}} \rb =\int_{0}^{\infty}\prod_{\V{s}}dp_{\V{s}}\, \delta\lb \sum_{ \V{s}}p_{\V{s}} -1  \rb 
\exp \lbb 
-\frac{1}{2E}\sum_{\mu=1}^{M} \lb \V{v}^{\mu}\cdot \lb \V{p}-\hat{\V{p}} \rb  \rb^{2} 
+\Gamma S(\V{p})
\rbb.
\Leq{volume}
\ee
For $\Gamma =0$ we recover the `unbiased' measure in \Req{measure}, while the limit $\Gamma\to\infty$ retains the ME distribution only.

%%%%%%%%%%%%%%%%%%%%%%%%%%%%%%%%%%%%%%%%%
\subsection{Randomization of observables}

Hereafter we assume that the $M$ constraints $\{ \V{v}^{\mu}\}_{\mu=1}^{M}$ are randomly and independently chosen from the following Gaussian distribution over the space of $2^N$--dimensional vectors $\V{v}$: 
\be
P(\V{v})=\prod_{\V{s}}\frac{1}{\sqrt{2\pi}}e^{-\frac{1}{2}v^{2}_{\V{s}}}.
\Leq{prior}
\ee
As mentioned in the introduction we do not pretend that this assumption is realistic. Indeed in practice, observables are often chosen to be low-order moments of the target distribution such as magnetizations or pairwise spin correlations, which is quite distinct from the Gaussian distribution above. 

In contrast to the constraints $\{\V{v}^{\mu}\}_{\mu}$, which are randomly drawn, we do not choose any particular statistical ensemble for the target distribution $\hat{\V{p}}$; The only properties of $\hat{\V{p}}$ we need for the analysis is the existence of the entropy curve discussed above. We stress that these constraints $\{\V{v}^{\mu}\}_{\mu}$, appearing in the measure $\rho$ over the distribution space, are quenched random variables, drawn once for all. 

Since each constraint contributing a multiplicative factor $e^{-(\V{v}^{\mu}\cdot (\V{p}-\hat{\V{p}} ) )^2/2E}$ to the volume $V$ in \Req{volume}, we expect that the logarithm of this volume will be self-averaging  in the large--$M$ limit. We will therefore calculate the average value of $\log V$ over the constraints, using the replica method. 
%\be
%\lsb F \rsb=\lim_{n\to 0}\frac{1}{n}\log \lsb V^{n}\rsb.
%\ee
%We mainly investigate $\lsb F\rsb$ under the replica symmetric (RS) ansatz, which is shown to be reasonable by examining the stability later. 

%%%%%%%%%%%%%%%%%%%%%%%%%%%%%%%%%%%%%%%%%
%%%%%%%%%%%%%%%%%%%%%%%%%%%%%%%%%%%%%%%%%
\section{Overview of results}\Lsec{overview}

We hereafter report the main outcomes of our replica calculation in an informal way. All technical details are postponed to \Rsec{analysis}. 

%%%%%%%%%%%%%%%%%%%%%%%%%%%%%%%%%%%%%%%%%
\subsection{Order parameters: interpretation and scaling with $N$}\Lsec{Order parameters}

Given the set of constraints $\{\V{v}^\mu\}_{\mu=1}^M$ and the target distribution $\hat{\V{p}}$ we may draw a schematic picture of the version space such as the one shown in \Rfig{space}. In addition to the target distribution, two distributions of interest in the version space are the ME distribution, $\V{p}^{\rm ME}$, and the center-of-mass distribution,
$\langle \V{p}\rangle$, where the angular brackets denote the average over the measure $\rho$:
\be
\langle \V{p}\rangle = \int d\V{p}\; \rho(\V{p})\; \V{p} \ .
\ee
The following quantities are useful to characterize the `distances' between the distributions in the version space, see \Rfig{space}:
\be
R(\{ \V{v}^{\mu} \}_{\mu},\hat{\V{p}})\equiv\sum_{\V{s}}\lb \Ave{p_{\V{s}}}-\hat{p}_{\V{s}} \rb^2\ ,\quad
D(\{ \V{v}^{\mu} \}_{\mu},\hat{\V{p}})\equiv \sum_{\V{s}} \lb \Ave{p_{\V{s}}^2}-\Ave{  p_{  \V{s}  }  }^2 \rb \ .
\ee
$R$ measures the $L_2$-squared distance between the target and the center-of-mass distributions. $D$ gives the size of the squared fluctuations of $\V{p}$ around the center of mass. 
We will also consider hereafter 
\be\Leq{defQ2}
Q(\{ \V{v}^{\mu} \}_{\mu},\hat{\V{p}})\equiv R+D = \sum_{\V{s}}\Ave{\lb p_{\V{s}}-\hat{p}_{\V{s}} \rb^2} \ , 
\ee
which measures the averaged square distance between $\V{p}$ and $\hat{\V{p}}$ (\Rfig{space}). 

\begin{figure}[htbp]
\begin{center}
\includegraphics[width=0.40\columnwidth]{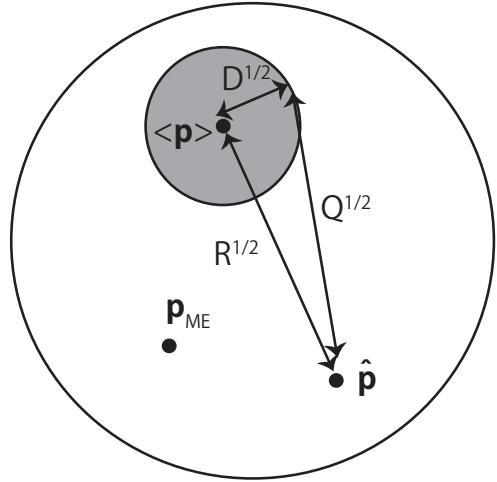}
\caption{Schematic view of the space of distributions. The large circle represents the version space of all admissible probability vectors, see \Req{constraint}, which depends on the constraints $\{\V{v}^\mu\}_{\mu=1}^M$. The shaded area represents the typical fluctuations of $\V{p}$ around the center-of-mass distribution $\Ave{\V{p}}$, of magnitude $\sqrt D$. The distance between the target distribution $\hat{\V{p}}$ and $\Ave{\V{p}}$ is $\sqrt{R}$, while $\sqrt{Q}$ is the square root of the averaged squared distance between $\V{p}$ and $\hat{\V{p}}$. These three order parameters measure the lengths of the sides of the rectangular triangle shown in the figure, see \Req{defQ2}. The ME distribution, $\V{p}^{\rm ME}$, lies inside the version space.}
\Lfig{space}
\end{center}
\end{figure}

$D,Q,R$ are intimately related to the statistics of the generalization error, that is, the error on the prediction of the average value of a new observable. Assume we have a measure $\rho$ over the version space defined from a set of constraints $\{\V{v}^\mu\}_{\mu=1}^M$. Let us now consider a new observable, $\V{v}'$. The error on the average value of this observable $\V{v}'$ computed with a distribution $\V{p}$ is simply given by
\be
\Delta (\V{p})=\V{v}'\cdot\lb \V{p}-\hat{\V{p}}\rb.
\ee
Suppose now that $\V{p}$ is randomly chosen according to measure $\rho$ and that the components $v'_{\V{s}}$ are independent and normal random variables, with zero means and unit variances. Let us denote the average over $\V{v}'$ by $\lsb \cdots \rsb_{\V{v}^{'}}$. It is easy to show that
\be
Q=\lsb \Ave{\Delta^2} \rsb_{\V{v}'}\ , \quad 
R=\lsb \Ave{\Delta}^2 \rsb_{\V{v}'}=
\lsb \Ave{\Delta}^2 \rsb_{\V{v}'}-\lsb \Ave{\Delta} \rsb_{\V{v}'}^2\ , \quad 
D=\lsb \Ave{\Delta^2} - \Ave{\Delta}^2 \rsb_{\V{v}'}.
\ee
Hence, $Q$ represents the averaged square error on a new observable, $R$ quantifies the observable-to-observable squared fluctuations of the error, and $D$ measures the distribution-to-distribution squared fluctuations of the error. The squared error on the prediction of the average value of the new observable, $Q$, is the sum of two contributions. The first one, $R$, is due to the choice of observable. The second one, $D$, reflects the distribution-to-distribution fluctuations with the version space measure. 

To understand the scaling of $D,Q,R$ with $N$ let us consider the simple case of no constraint at all, $M=0$, and no bias on the entropy, $\Gamma=0$. In this case, due to the permutation symmetry over the $2^N$ configurations, all the configuration probabilities $p_{\V{s}}$ obey the same marginal distribution, $\rho_{\V{s}}(p_{\V{s}})$, see \Req{marginal}. It is easy to convince oneself, that $\rho_{\V{s}}$ becomes a pure exponential when $N\gg 1$, with average value $\langle p_{\V{s}}\rangle=2^{-N}$. Indeed, the volume in \Req{volume1} is given by
\be
V=\int_{0}^{\infty}\prod_{\V{s}}dp_{\V{s}}~\delta\lb \sum_{\V{s}}p_{\V{s}}-1 \rb
=\int_{-i\infty}^{i\infty} d\Lambda~e^{\Lambda}  \int_{0}^{\infty} \prod_{\V{s}} \lb dp_{\V{s}}~e^{ -\Lambda p_{\V{s}} } \rb=\int_{-i\infty}^{i\infty} d\Lambda~\frac{e^\Lambda}{\Lambda^{2^{N}}}\ ,
\Leq{volume-noconst}
\ee
where we have used an integral representation of the Dirac delta function. This can be directly integrated but we here use the saddle-point method, valid for large $N$ for later convenience. The saddle point is located at
\be
\Lambda=2^N \ ,
\Leq{Lambda-noconst}
\ee
giving $\log V \doteq 2^N(1- N\log 2)$, which agrees with the leading behaviour of the volume of the $2^N$--dimensional simplex, $V=1/[(2^N)!]$. In addition we see from \Req{marginal} and \Req{volume-noconst} that $\rho_{\V{s}}$ is an exponential distribution with mean value $1/\Lambda=2^{-N}$, as announced above. Hence,
\be
 D\doteq e^{Nd}\ ,\ Q\doteq e^{Nq}\ ,\  R\doteq e^{Nr}\ ,\  \Lambda\doteq e^{N\lambda}\ ,
\ee
with $q=r=-\ell_2$, see \Req{lk}, and $d=-\lambda=\log 2$. Our calculation with the replica method shows that, in the presence of constraints, {\em i.e.} for $M\ge 1$, the exponential--in--$N$ scaling of $D,Q,R,\Lambda$ will still hold after averaging over the constraints, even if a bias $\Gamma$ over the entropy is imposed. Note that the rates $d,q,r,\lambda$ will then depend on $M$ and $\Gamma$.

%The external parameters $M,E$ and $\Gamma$ also have the counterparts in the exponent
%\be
%\Lambda\doteq e^{N\lambda} , M\doteq e^{N\alpha},\,\, E\doteq e^{N\epsilon},\,\, \Gamma \doteq e^{N\gamma}.
%\ee

%%%%%%%%%%%%%%%%%%%%%%%%%%%%%%%%%%%%%%%%%
\subsection{Description of the learning process}
\Lsec{learning edge}

We first focus on the effect of increasing the number of measured observables, $M$, and set the entropic bias $\Gamma$ to zero. The effect on non-zero biases will be reported in \Rsec{role-ME}. We assume in \Rsec{les1} and \Rsec{les2} that the tolerance $E$ is negligible; the dependence of the results on the value of $E$ is exposed in \Rsec{les3}.

%%%%%%%%%%%%%%%%%%%%%%%%%%%%%%%%%%%%%%%%%
\subsubsection{The learning edge}
\Lsec{les1}

One major result of the replica calculation is that the marginal measure over the single-configuration probabilities, $\rho_{\V{s}}$, \Req{marginal}, may have two distinct behaviours, sketched in \Rfig{schematic rho}:
\begin{itemize}
\item The marginal measure over the probability $p_{\V{s}}$ of a configuration $\V{s}$ having a {\em large} value $\hat{p}_{\V{s}}$ in the target distribution is concentrated at a value close to this target probability:
\be
\rho_{\V{s}}(p_{\V{s}})\approx A\, e^{-B\, (p_{\V{s}}-\hat{p}_{\V{s}} -\delta_{\V{s}}   )^2/2}\  ,
\Leq{dist-large}
\ee
where the values of $A$ and $B$ depend on the parameters $M,E$, and will be specified later. The corresponding curve of $\rho_{\V{s}}$ is shown in \Rfig{schematic rho}, left. The shift $\delta_{\V{s}}$ is small compared to the target probability $\hat{p}_{\V{s}}$, as we will see in \Rsec{analysis}. In other words, the configuration probability $\hat{p}_{\V{s}}$ is correctly `learned' from the values of the constraints.

\item The marginal measure over the probability $p_{\V{s}}$ of a configuration $\V{s}$ having a {\em small} value $\hat{p}_{\V{s}}$ in the target distribution is a decaying exponential: 
\be
\rho_{\V{s}}(p_{\V{s}})\approx \frac{1}{A'}\, e^{-p_{\V{s}}/A'} \ .
\Leq{dist-small}
\ee
The corresponding curve of $\rho_{\V{s}}$ is shown in \Rfig{schematic rho}, right.  The average value $A'$ of the configuration probability $p_{\V{s}}$ does not depend on $\hat{p}_{\V{s}}$ in the dominant scaling with $N$. $A'$ is a function of the parameters $M,E$ only, and will be specified later. In other words, the configuration probability is not at all inferred.
\end{itemize}

%%%%%%%%%%%%%%%%%%%%%%%%%%%
\begin{figure}[htbp]
\begin{center}
\includegraphics[width=0.9\columnwidth]{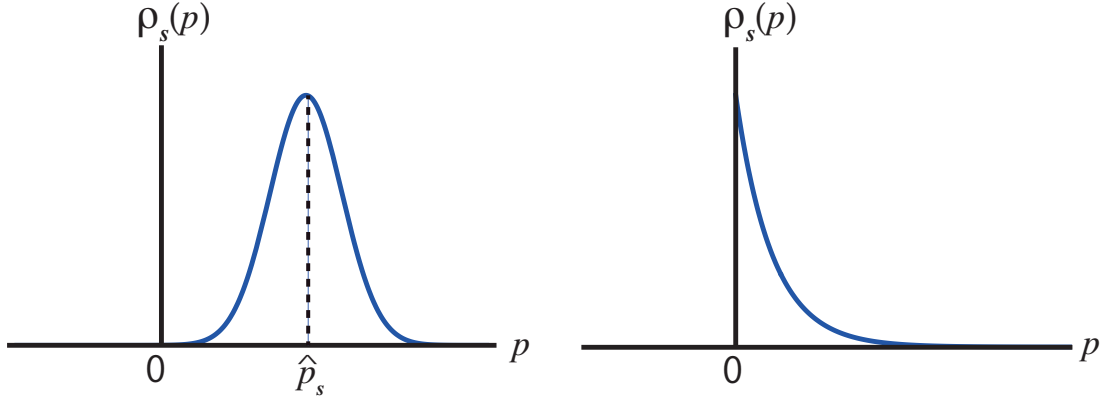}
\caption{Schematic pictures of the marginal measure $\rho_{\V{s}}(p_{\V{s}})$ for large (left) and small (right) configuration probabilities $\hat{p}_{\V{s}}$. }
\Lfig{schematic rho}
\end{center}
\end{figure}
%%%%%%%%%%%%%%%%%%%%%%%%%%%

The concepts of {\em large} and {\em small} target probabilities are defined as follows:
\be\Leq{le}
\hat p_{\V{s}}\doteq e^{-N\,\omega_{\V{s}}} \ \text{ is large if} \ \omega_{\V{s}} <\hat \omega\ , 
\ \text{ and is small if} \ \omega_{\V{s}} >\hat \omega\ . 
\ee
The boundary $\hat \omega$ is hereafter referred to as the {\em learning edge}, as it separates the set of probabilities $p_{\V{s}}$  into the ones which can be correctly learned and the ones which remain essentially unknown, despite the knowledge of the observable values. Its value depends on the parameters $M$ and $E$. The representative curve of the learning edge $\hat \omega$ is an increasing function of the rate
\be
\alpha = \frac{\log M}N\ ,
\ee
and is shown in \Rfig{order parameters-analytics-1a} for the ISM and a negligible tolerance $E$. We find qualitative changes of the curve $\hat\omega(\alpha)$, taking place at critical values of the ratio $\alpha$ and corresponding to the onset of phase transitions in the learning process.  

%%%%%%%%%%%%%%%%%%%%%%%%%%%
\begin{figure}[htbp]
\begin{center}
\includegraphics[width=0.45\columnwidth]{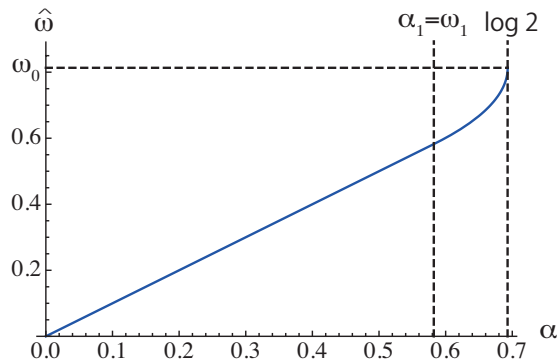}
\caption{Learning edge $\hat \omega$  as a function of $\alpha$ for the ISM \NReq{ISM} with $H=0.5$ and for small $E$.}
\Lfig{order parameters-analytics-1a}
\end{center}
\end{figure}
%%%%%%%%%%%%%%%%%%%%%%%%%%%

%%%%%%%%%%%%%%%%%%%%%%%%%%%%%%%%%%%%%%%%%
\subsubsection{Phase transitions and typical distances between distributions}
\Lsec{les2}

Given the value of the learning edge $\hat\omega$ we define, according to \Req{le}, the sets of configurations $\V{s}$ having large and small probabilities as, respectively, $L$ and $S$. Our replica calculation shows that the order parameters $Q$ and $R$ are given by, to the dominant order in $N$, 
\be
Q\doteq R\doteq \sum_{\V{s}\in S}\hat{p}_{\V{s}}^2.
\Leq{square-sum}
\ee
If $M$ is small, so is the learning edge $\hat{\omega}$, most configurations are found in $S$. This implies $\sum_{\V{s}\in S}\hat{p}_{\V{s}}^2 \doteq \sum_{\V{s}}\hat{p}_{\V{s}}^2=e^{-N\ell_2}$. As $M$ increases, the learning edge grows, and reaches $\hat{\omega}=\omega_{2}$. For larger values of $M$, the dominant point $\omega=\omega_{2}$ is now not in $\V{s}\in S$ and hence the rates $q$ and $r$ change  from $-\ell_2$. This leads non-analyticities in $Q$ and $R$, and the transition point is at $\hat{\omega}=\omega_{2}$. Similarly, a switch of the dominant term in the normalization condition $1=\sum_{\V{s}\in S}\hat{p}_{\V{s}}+\sum_{\V{s}\in L}\hat{p}_{\V{s}}$ from the $S$ to the $L$ occurs at $\hat{\omega}=\omega_{1}$. The Lagrange multiplier $\Lambda$ is non analytic at this point, which defines a second phase transition. 

Hence, we have three distinct phases, separated by specific values of the learning edge. We label the phases by `I' when $\hat{\omega} < \omega_2$, `II' when $\omega_{2} \leq \hat{\omega} < \omega_1$, and `III' when $\omega_{1} \leq \hat{\omega}$. The value of the ratio $\alpha$ corresponding to the transition point $\hat{\omega}=\omega_2$ is denoted as $\alpha_2$, and the one corresponding to $\hat{\omega}=\omega_1$ is called $\alpha_1$. In phases I and II with small $E$, $\hat{\omega}$ turns out to be equal to $\alpha$. Thus $\alpha_2=\omega_2$ and $\alpha_1=\omega_1$.  

To get insights about the typical distances between the distributions of interest in the version space we plot the order parameters $q,r$ and $d$ of the ISM in \Rfig{order parameters-analytics-1b}. Those order parameters continuously change at the transition points. At $\alpha=\log 2$, the learning edge becomes equal to $\omega_{0}$; $q$ and $r$ reach the same value as $d$. The agreement between $r$ and $d$, the distance of the center of mass to the target distribution and the typical fluctuations, implies that the volume largely shrinks and scales with $N$ in a different manner at this point. Our calculation, restricted to the leading order in the volume, is not informative for larger ratios, {\em i.e.} $\alpha>\log 2$.

%%%%%%%%%%%%%%%%%%%%%%%%%%
\begin{figure}[htbp]
\begin{center}
\includegraphics[width=0.45\columnwidth]{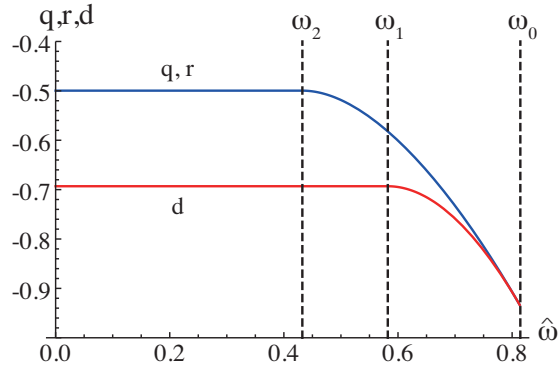}
\caption{Order parameters $q,r,d$ as functions of the learning edge $\hat{\omega}$ for the ISM \NReq{ISM} with $H=0.5$ and for small $E$. The critical values of $\omega$ are $\omega_0\approx0.81$, $\omega_1\approx0.58$, and $\omega_2\approx0.43$, and the corresponding entropies are $\sigma_0=\log 2\approx 0.69$, $\sigma_1\approx0.58$, and $\sigma_2\approx 0.37$, respectively. }
\Lfig{order parameters-analytics-1b}
\end{center}
\end{figure}
%%%%%%%%%%%%%%%%%%%%%%%%%%%

%%%%%%%%%%%%%%%%%%%%%%%%%%%%%%%%%%%%%%%%%
\subsubsection{Effect of the tolerance $E$}
\Lsec{les3}

Let us now consider the case of a non-negligible tolerance $E$.  We give the corresponding phase diagram in \Rfig{PD-quench-1}. It is natural to scale the tolerance as $E\doteq e^{N\epsilon}$. As $\epsilon$ grows from very negative values, we expect that the quality of the inference becomes worse, as the version space include more distributions, which do not exactly reproduce the target values of the observables. Indeed, for large values of the tolerance $E$, there emerge new phases where the learning edge decreases as $\epsilon$ grows. We call these phases with large tolerance ${\rm I}_{{\rm LT}}$, ${\rm II}_{{\rm LT}}$, and ${\rm III}_{{\rm LT}}$, in agreement with the denomination chosen above based on the different ranges of possible values for the learning edge.

%%%%%%%%%%%%%%%%%%%%%%%%%%%
\begin{figure}[htbp]
\begin{center}
\includegraphics[width=0.48\columnwidth]{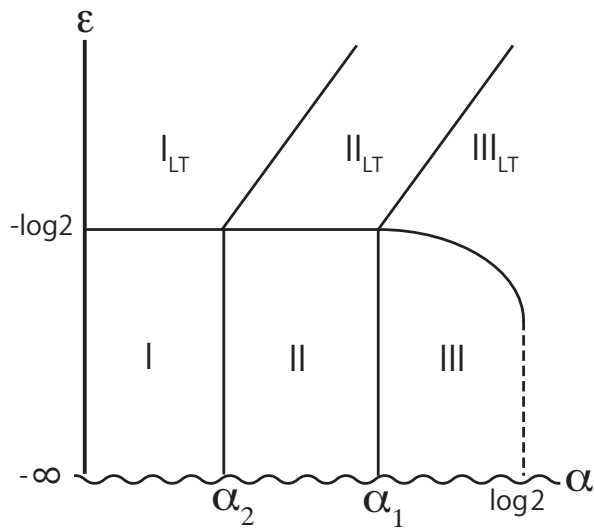}
\caption{Phase diagram in the $\alpha$-$\epsilon$ plane in the absence of any entropic bias.}
\Lfig{PD-quench-1}
\end{center}
\end{figure}
%%%%%%%%%%%%%%%%%%%%%%%%%%%

The transition from small to large tolerances takes place at $E \doteq D$. This result is easy to interpret: deviations in the values of observables from the target ones smaller than the scale of the intrinsic fluctuations $D$ will be masked by those fluctuations, and cannot degrade the inference performances. The results shown in \Rsec{les1} and \Rsec{les2} are therefore valid as long as $\epsilon < d$.

In the present study, we do not consider any measurement error: the measured values of the observables correspond exactly to the projection of the observable vectors onto the target distribution. If measurement errors were considered, the role of the tolerance parameter could possibly change; an appropriate amount of tolerance $E$, of the order of the measurement error, would lead to better inference by avoiding overfitting.

%%%%%%%%%%%%%%%%%%%%%%%%%%%%%%%%%%%%%%%%%
\subsection{Effects of the entropic bias}
\Lsec{role-ME}

We now consider the effect of an entropic bias $\Gamma >0$ on the learning performances. As $\Gamma$ grows, the measure $\rho$ gives more and more weight to the distributions $\V{p}$ with large entropies. In the $\Gamma \to \infty$ limit $\rho$ singles out the ME distribution. The $\Gamma$-dependence of the quantities of interest, in particular, the learning edge, is of key importance. Similarly to the other parameters we assume that the entropic bias scales $\Gamma=e^{N\gamma}$. 

We first consider the case of negligible $E\approx 0$. Our replica calculation shows that the order parameters $Q,R$ and the learning edge $\hat\omega$ do not depend at all on the value $\Gamma \geq 0$. An immediate consequence is that the ME distribution is not closer to the target distribution than any randomly picked up distribution (with the unbiased measure) in the version space.  
%has no superiority at all to other distributions in the version space. However, we believe that this reluctant (though nontrivial) outcome is due to our assumption that the observables are randomly independently drawn from \Req{prior}. Relaxing this assumption is an important future work and currently under way. The result will be reported elsewhere.  
Nevertheless the presence of an entropic bias affects the fluctuations of $\V{p}$, measured by the order parameter $D$. For $\alpha<\alpha_1$, our replica calculation shows that $D\doteq 2^{-N}$ for small $\Gamma$, while $D \doteq \Gamma^{-1}$ for large $\Gamma$. The transition between these two scalings takes place at $\gamma_c=\log 2$. The same transition for $D$ is observed when $\alpha>\alpha_1$, though the transition point $\gamma_c$ now depends on $\alpha$. The order parameter $\Lambda$ shows non-analyticities at those transition points, contrary to $Q$, $R$, and the learning edge, which remain unaffected as mentioned above. The non-analyticities in $D$ and $\Lambda$ discriminate the phases for $\gamma \leq \gamma_c$ from the ones for $\gamma>\gamma_c$; we denote the phases with large entropic bias by ${\rm I}_{{\rm ME}}, {\rm II}_{{\rm ME}},$ and ${\rm III}_{{\rm ME}}$, in agreement with the denomination based on the ranges of values of the learning edge. The corresponding phase diagram is shown in \Rfig{PD-quench}, left. On the right panel of the same figure, the order parameter $d$ is plotted against $\gamma$ for the ISM with $H=0.5$ and $\alpha< \alpha_1$ as an illustration. 
%%%%%%%%%%%%%%%%%%%%%%%%%%%
\begin{figure}[htbp]
\begin{center}
\includegraphics[width=0.4\columnwidth]{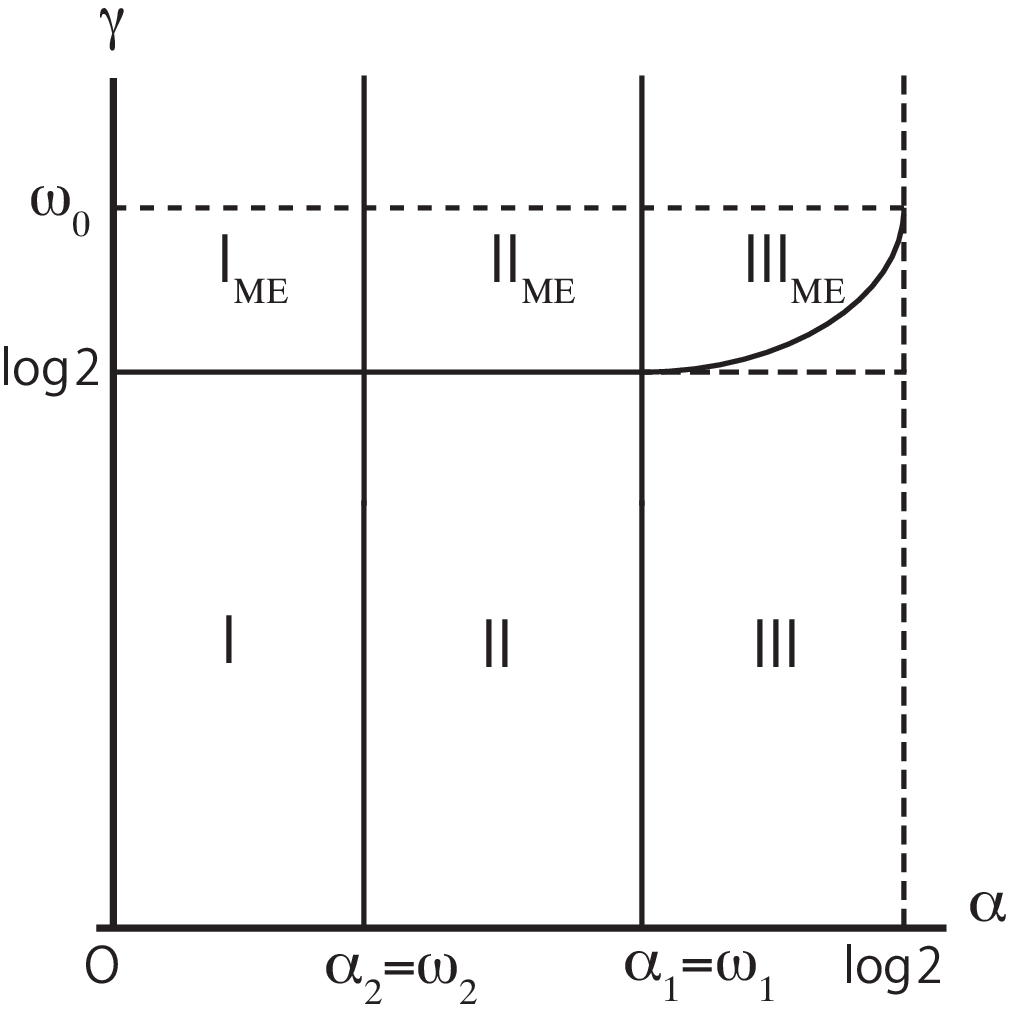}
\hskip 1cm
\includegraphics[width=0.5\columnwidth]{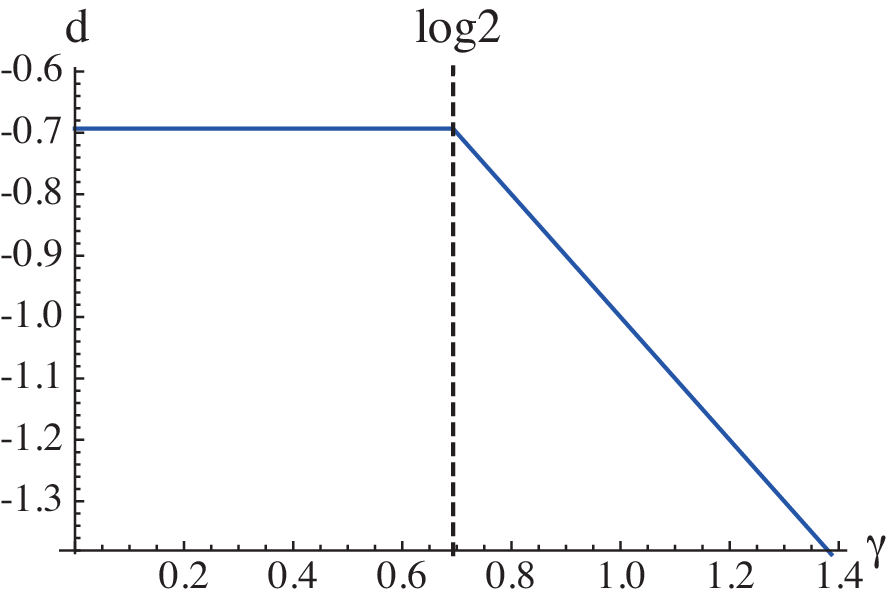}
\caption{Left: Phase diagram in the $\alpha$-$\gamma$ plane for very small tolerance $E$. The fluctuation order parameter $D\doteq e^{Nd}$ depends on $\gamma$ in the phases with large entropic biases, ${\rm I}_{{\rm ME}}, {\rm II}_{{\rm ME}},$ and ${\rm III}_{{\rm ME}}$, while it keeps the same value as for $\Gamma=0$ in ${\rm I}, {\rm II},$ and ${\rm III}$. Right: order parameter $d$ as a function of $\gamma$ for the ISM and $\alpha<\alpha_1$. A non-analyticity appears at $\gamma=\gamma_c(=\log 2)$, which signals the onset of the phase with large entropic bias for $\gamma>\gamma_c$.  }
\Lfig{PD-quench}
\end{center}
\end{figure}
%%%%%%%%%%%%%%%%%%%%%%%%%%%

The phases above change when the value of the tolerance $E\doteq e^{N\epsilon}$ is not negligible any longer. As in the $\Gamma=0$ case the tolerance matters if $E>D$. As the fluctuations scale as $D \doteq \Gamma^{-1}$ for large $\Gamma$ with $\alpha<\alpha_1$, this implies that a regime takes place for $ \gamma > -\epsilon$. This large tolerance, large entropic bias regime shows unusual properties. For instance the learning edge becomes smaller and smaller, that is, the learning performances become worse and worse, as $\Gamma$ grows, in contrast to the case of negligible $E\approx 0$ where the learning edge has no dependence on $\Gamma$. This seemingly-strange behaviour is actually expected. We see from \Req{volume} that, for very large $\Gamma$ and finite fixed $E$, the constraints become irrelevant. Hence, in this limit, the measure concentrates around the uniform distribution $p^{\rm ME}_{\V{s}}=1/2^N$, maximizing the entropy irrespective of the constraints. To get an meaningful ME distribution, the tolerance $E$ must be enough small compared to the fluctuations $D$ governed by $\Gamma$ in the large $\Gamma$ limit. In conclusion, this large tolerance, large entropic bias regime is irrelevant and will not be studied further.

%%%%%%%%%%%%%%%%%%%%%%%%%%%%%%%%%%%%%%%%%
%%%%%%%%%%%%%%%%%%%%%%%%%%%%%%%%%%%%%%%%%
\section{Analytical calculations}
\Lsec{analysis}
In this section, we present our replica calculation of the average value of the logarithm of the volume $V$ over the constraints. We introduce the notations
\be
 \Tr{\V{p}} (\cdots)\equiv
 \int_{0}^{\infty}\prod_{\V{s}}dp_{\V{s}}\delta\lb \sum_{\V{s}}p_{\V{s}}-1 \rb (\cdots),
\ee
and $Dz\equiv dz e^{-\frac{1}{2}z^2}/\sqrt{2\pi}$. We also write undefined integrals as a shorthand notation for the domains of integration $[-\infty,\infty]$ or $[-i\infty,i\infty]$.

%%%%%%%%%%%%%%%%%%%%%%%%%%%%%%%%%%%%%%%%%
\subsection{Replica calculation of the average logarithm of the volume}

We start by unraveling the squared terms in the exponent of volume \NReq{volume} through auxiliary Gaussian integrations, or the so-called Hubbard-Stratonovich transformations:
\be
V=\int \prod_{\mu=1}^{M} Dz_{\mu} \Tr{\V{p}} 
\exp\lbb i\sum_{\mu}\sum_{\V{s}}\frac{ z_{\mu} }{ \sqrt{E} } v^{\mu}_{\V{s}}(p_{\V{s}}-\hat{p}_{\V{s}}) + \Gamma S(\V{p}) \rbb.
\Leq{HST}
\ee
According to the replica method we calculate the $n$th moment of the volume of $n \in \mN$
\be
&&
\lsb V^n \rsb=
\prod_{a=1}^{n}\lb  \int \prod_{\mu} Dz_{\mu}^{a} \Tr{ \V{p}^a }  \rb
 \lsb 
\exp\lbb i\sum_{\mu}\sum_{\V{s}}\sum_a \frac{ z^a_{\mu} }{ \sqrt{E} } v^{\mu}_{\V{s}}(p^a_{\V{s}}-\hat
{p}_{\V{s}}) +\Gamma\sum_a  S(\V{p}^a) \rbb
 \rsb
\no \\
&&=
\prod_{a=1}^{n}\lb  \int \prod_{\mu} Dz_{\mu}^{a} \Tr{ \V{p}^a }  \rb
\exp
 \lbb 
   -\sum_{\mu} \sum_{a,b} \frac{ z^a_{\mu}z^b_{\mu} }{ 2E } Q_{ab}
   +\Gamma\sum_a  S(\V{p}^a) 
  \rbb
\no \\
&&= \prod_{a=1}^{n}\Tr{ \V{p}^a } 
\exp
\lbb
-\frac{M}{2}\log \det \lb 1 +\frac{Q}{E}\rb+ \Gamma \sum_a S(\V{p}^a) 
\rbb,
\Leq{replica-1}
\ee
where the square brackets denote the average over the observables $\{ \V{v}^{\mu}\}_{\mu=1}^M$ as in \Rsec{Order parameters}, and we define
\be
Q_{ab}=\sum_{\V{s}}(p^{a}_{\V{s}}-\hat{p}_{\V{s}})(p^{b}_{\V{s}}-\hat{p}_{\V{s}}).
\ee
To perform the integrations $\Tr{\V{p}^{a}}$, we make use of the following identity
\be
&&1=\int \prod_{a\leq b} dQ_{ab}
\prod_{a\leq b} 
\delta \lb
Q_{ab} -\sum_{\V{s}}(p^{a}_{\V{s}}-\hat{p}_{\V{s}})(p^{b}_{\V{s}}-\hat{p}_{\V{s}})
\rb
\no \\
&&=
C \int \prod_{a\leq b} dQ_{ab}dQ'_{ab} 
\exp
 \lbb
  \frac{1}{2}\sum_{a\leq b}Q_{ab}Q'_{ab}
  -\frac{1}{2}\sum_{a\leq b}\sum_{\V{s}}Q'_{ab}(p^{a}_{\V{s}}-\hat{p}_{\V{s}})(p^{b}_{\V{s}}-\hat{p}_{\V{s}})
 \rbb,
\ee
where the normalization constant $C$ is irrelevant and will be discarded hereafter. Similarly, we rewrite $\Tr{\V{p}^{a}}$ as
\be
\Tr{ \V{p}^a }=\int d\Lambda_a \int_{0}^{\infty} \prod_{\V{s}}dp^{a}_{\V{s}}\; e^{-\Lambda_a\sum_{\V{s}}(p^{a}_{\V{s}}  -\hat{p}_{\V{s}})  },
\ee
where we used the normalization identity $1=\sum_{\V{s}}\hat{p}_{\V{s}}$. We then obtain
\be
&&[V^n]=
\int \prod_{a\leq b} dQ_{ab}dQ'_{ab}\int \prod_{a}d\Lambda_a
 \int_{0}^{\infty}\prod_{a}\prod_{\V{s}} dp^a_{\V{s}}
\no \\ &&
\times 
\exp 
 \lb
-\sum_{a}\Lambda_a\sum_{\V{s}}(p^{a}_{\V{s}}-\hat{p}_{\V{s}})-\frac{1}{2}\sum_{a\leq b}\sum_{\V{s}}Q'_{ab}(p^{a}_{\V{s}}-\hat{p}_{\V{s}})(p^{b}_{\V{s}}-\hat{p}_{\V{s}})
+\Gamma \sum_{a}S(\V{p}^a)
 \rb
 \no \\
&&
\times
\exp
\Biggl\{
 \frac{1}{2}\sum_{a\leq b}Q_{ab}Q'_{ab}
  -\frac{M}{2}\log \det \lb 1 +\frac{Q}{E}\rb 
\Biggr\}.
\ee
The logarithm of $[V^n]$ can be approximated by the saddle-point value of
\be
\phi(n)\equiv \frac{1}{2}\sum_{a\leq b}Q_{ab}Q'_{ab}
  -\frac{M}{2}\log \det \lb 1 +\frac{Q}{E}\rb 
+\sum_{\V{s}} \log\Theta_{\V{s}},
\ee
over the matrices $Q$ and $Q'$, and where
\be
\Theta_{\V{s}}=\int_{0}^{\infty}\prod_{a}dp_{\V{s}}^{a}
\exp 
 \lb
-\sum_{a}\Lambda_a(p^{a}_{\V{s}}-\hat{p}_{\V{s}})
-\frac{1}{2}\sum_{a\leq b}Q'_{ab}(p^{a}_{\V{s}}-\hat{p}_{\V{s}})(p^{b}_{\V{s}}-\hat{p}_{\V{s}})
-\Gamma \sum_{a}p^a_{\V{s}}\log p^a_{\V{s}}
 \rb.
 \Leq{Theta}
\ee

%%%%%%%%%%%%%%%%%%%%%%%%%%%%%%%%%%%%%%%%%
%\subsubsection{Replica Symmetry}

We now assume that the order parameter matrices are invariant under permutation symmetry of the replica indices, and write
\be
Q_{ab}=R+(Q-R)\delta_{ab},\,\,
Q'_{ab}=-2R'+(Q'+R')\delta_{ab}, \,\, 
\Lambda_a=\Lambda,
\Leq{RS ansatz}
\ee
where $\delta_{ab}$ is the Kronecker delta. Thus
\be
&&\frac{1}{2}\sum_{a\leq b}Q_{ab}Q'_{ab}=\frac{1}{2}nQ(Q'-R') 
                                                   -\frac{1}{2}n(n-1)RR',
\\
&&
\log \det \lb 1+ \frac{Q}{E} \rb=
\log \lb1+ \frac{Q+(n-1)R)}{E} \rb +(n-1)\log \lb1+ \frac{Q-R}{E} \rb,
\ee
and,
\be
\frac{1}{2}\sum_{a\leq b}Q'_{ab}(p^{a}_{\V{s}}-\hat{p}_{\V{s}})(p^{b}_{\V{s}}-\hat{p}_{\V{s}})
=\frac{1}{2}Q'\sum_{a} (p_{\V{s}}^{a}-\hat{p}_{\V{s}})^2
-\frac{1}{2}R'\lb \sum_a (p_{\V{s}}^{a}-\hat{p}_{\V{s}}) \rb^2.
\ee
Using the Hubbard-Stratonovich transformation again \Req{Theta} becomes
\be
\Theta_{\V{s}}=\int Dz 
\int_{0}^{\infty}\prod_{a}dp_{\V{s}}^{a} 
e^{ 
-\lb \Lambda -z \sqrt{R'} \rb \lb p^{a}_{\V{s}}-\hat{p}_{\V{s}} \rb
-\frac{1}{2}Q'(p^{a}_{\V{s}}-\hat{p}_{\V{s}})^2
-\Gamma p^a_{\V{s}}\log p^a_{\V{s}}
 }
 \equiv \int Dz X_{\V{s}}^n.
\ee
Hence,
\be
&&\phi(n)=\frac{1}{2}nQ(Q'-R')-\frac{1}{2}n(n-1)RR'
\no \\
&&
-\frac{M}{2}\lbb \log \lb1+ \frac{Q+(n-1)R)}{E} \rb +(n-1)\log \lb1+ \frac{Q-R}{E} \rb \rbb
+\sum_{\V{s}} \log \int Dz X_{\V{s}}^{n}.
\Leq{phi-general}
\ee
We finally obtain the expression for $\lsb \log V\rsb =\lim_{n \to 0}\phi(n)/n$, 
\be
\lsb \log V \rsb=\frac{1}{2}Q(Q'-R')+\frac{1}{2}RR'
-\frac{M}{2}\lbb \frac{R}{E +Q-R}+\log \lb1+ \frac{Q-R}{E} \rb \rbb
+\sum_{\V{s}} \int Dz  \log X_{\V{s}},
\Leq{F-quench}
\ee
where
\be
X_{\V{s}}=\int_{0}^{\infty}~dp~e^{
-\lb \Lambda-z\sqrt{R'} \rb \lb p-\hat{p}_{\V{s}} \rb
-\frac{1}{2}Q'(p-\hat{p}_{\V{s}})^2
-\Gamma p \log p
}.
\Leq{X-general}
\ee

The integration over $p$ in \Req{X-general} has to be done with care. We assume that the conjugated order parameters, $Q'$ and $R'$, obey an exponential scaling with $N$,
\be
Q'\doteq e^{Nq'},\,\, R'\doteq e^{Nr'}.
\ee
Therefore, all the parameters appearing in the integration diverge or vanish when $N\to\infty$. We need to find out which order parameters diverge or not, consistently with the equations of state derived below. These procedures require involved case analyses. The outcome is that the equations of state are consistent, to the leading order in $N$, if the following conditions are met:
\be
0<\lambda < q'\leq 2 \lambda,\ 0<\lambda < r' \leq 2 \lambda,\ q=r<0, \ \epsilon< -\gamma \leq 0.
\Leq{magnitude}
\ee

%%%%%%%%%%%%%%%%%%%%%%%%%%%%%%%%%%%%%%%%%
\subsection{$\Gamma=0$ case}

In this case, the integral in \Req{X-general} can be evaluated in two different ways. The first one is more direct, but the latter one is useful to clarify the physical significance of the solutions and to treat the finite--$\Gamma$ case. 

%%%%%%%%%%%%%%%%%%%%%%%%%%%%%%%%%%%%%%%%%
\subsubsection{Direct integration and expansion}
\Lsec{direct integration}
The first strategy to evaluate $X_{\V{s}}$ is to directly integrate \Req{X-general} in $\Gamma=0$. The result is
\be
&&
X_{\V{s}}=
\int_{0}^{\infty}~dp~e^{
-\lb \Lambda-z\sqrt{R'} \rb \lb p-\hat{p}_{\V{s}} \rb
-\frac{1}{2}Q'(p-\hat{p}_{\V{s}})^2
}
=\sqrt{ \frac{ 2\pi   }{ Q' } }
e^{ \frac{1}{2}\frac{(\Lambda-z\sqrt{R'})^2}{ Q'  } }
H\lb y_{\V{s}}  -z\sqrt{ \frac{ R'  }{ Q' } }\rb,
\Leq{X-quench}
\ee
where we put $y_{\V{s}}=(\Lambda-\hat{p}_{\V{s}}Q')/\sqrt{Q'}$ and we define the complementary error function
\be
H(y)=\int_{y}^{\infty}Dz.
\ee
Thus
\be
&&\lsb \log V \rsb=\frac{1}{2}Q(Q'-R')+\frac{1}{2}RR'
-\frac{M}{2}\lbb \frac{R}{E +Q-R}+\log \lb1+ \frac{Q-R}{E} \rb \rbb
\no \\
&&
+2^{N}\lb \frac{ \Lambda^2+R' }{ 2Q' }+\frac{1}{2}\log 2\pi-\frac{1}{2}\log Q' \rb
+\sum_{\V{s}} \int Dz  \log H\lb y_{\V{s}}  -z\sqrt{ \frac{ R'  }{ Q' } }\rb.
\ee
The definition of $y_{\V{s}}$ allows us to introduce the learning edge in a natural way. Under assumption \NReq{magnitude}, $y_{\V{s}}=(\Lambda-\hat{p}_{\V{s}}Q')/\sqrt{Q'}$ diverges. If the target probability $\hat{p}_{\V{s}}\doteq e^{-N\omega_{\V{s}}}$ is small enough, {\em i.e.} if $\omega_{\V{s}}>q'-\lambda$, the dominant term in $y_{\V{s}}$ is $\Lambda/\sqrt{Q'}$, which goes to $+\infty$ in the thermodynamic limit. On the contrary we find  $y_{\V{s}}\to -\infty$ if $\omega_{\V{s}}<q'-\lambda$. This sharp difference defines the learning edge and the corresponding entropy as
\be
\hat{\omega}=q'-\lambda,
~\hat{\sigma}=\sigma(\hat{\omega})
\Leq{learning edge}
\ee
Furthermore, based on assumption \NReq{magnitude}, $|y_{\V{s}}|$ increases faster with $N\gg 1$ than $\sqrt{ R'/Q'  }$, entailing that the argument of the complementary error function $H$ in \Req{X-quench} is dominated by $y_{\V{s}}$. Thus, the complementary error function can be replaced with its asymptotic behavior, depending on the sign of $y_{\V{s}}$,
\be
H(y) \to 
\Biggl\{ 
\begin{array}{cl}
\frac{1}{\sqrt{2\pi}}\frac{1}{y}e^{-\frac{1}{2}y^2}, &  (y\to \infty) 
\\
1,  &  (y\to -\infty) %%+\frac{1}{\sqrt{2\pi}}\frac{1}{y}e^{-\frac{1}{2}y^2}
\end{array}
.
\Leq{H-limit}
\ee
Using this, we get
\be
&&
\sum_{ \V{s} }\int Dz \log H\lb y_{\V{s}}  -z\sqrt{ \frac{ R'  }{ Q' } }\rb
\no \\
&&
=
\int Dz
\sum_{\V{s} \in S}
\lbb
 -\frac{1}{2}\lb y_{\V{s}} -z\sqrt{ \frac{ R'  }{ Q' } }\rb^2
 -\frac{1}{2}\log 2\pi 
 -\log \lb y_{\V{s}}  -z\sqrt{ \frac{ R'  }{ Q' } }\rb
\rbb
+\sum_{\V{s} \in L}0.
\ee
The logarithmic term in the above formula can be expanded according to \Req{magnitude} as
\be
\log \lb y_{\V{s}}  -z\sqrt{ \frac{ R'  }{ Q' } }\rb
\approx
\log \Lambda-\frac{1}{2}\log Q'-\frac{\hat{p}_{\V{s}}Q'+\sqrt{R'}z}{\Lambda}
-\frac{1}{2}\frac{(\hat{p}_{\V{s}} Q'+\sqrt{R'}z)^2  }{\Lambda^2}.
\ee
Linear terms with respect to $z$ vanish upon Gaussian integration. We are left with
\be
&&
[\log V] \approx \frac{1}{2}Q(Q'-R')+\frac{1}{2}RR'
-\frac{M}{2}\lbb \frac{R}{E +Q-R}+\log \lb1+ \frac{Q-R}{E} \rb \rbb
\no \\
&&
+e^{ N\hat{\sigma} }\lb \frac{ \Lambda^2+R' }{ 2Q' }+\frac{1}{2}\log 2\pi-\frac{1}{2}\log Q' \rb
-2^{N}\lb \log \Lambda-\frac{1}{2}\frac{ R' }{\Lambda^2}\rb
\no \\
&&
+\lb \Lambda+\frac{ Q' }{\Lambda} \rb\sum_{ \V{s} \in S}\hat{p}_{\V{s}}
-\frac{1}{2}\lb Q'- \frac{ Q'^2 }{ \Lambda^2 } \rb\sum_{\V{s}\in S}\hat{p}_{\V{s}}^2,
\Leq{F-quench2}
\ee
where we have used
\be
2^{N}-\sum_{\V{s} \in S}1=\sum_{\V{s} \in L}1\doteq e^{N\hat{\sigma} },\
\sum_{\V{s} \in S}1\doteq 2^{N}.
\ee
These formulas are correct as long as $\hat{\omega} \leq \omega_0$.

%%%%%%%%%%%%%%%%%%%%%%%%%%%%%%%%%%%%%%%%%
\subsubsection{Saddle-point approximation}
\Lsec{saddle-point approx}
Our second method to evaluate $X_{\V{s}}$ in \Req{X-quench} is based on the saddle-point approximation. Writing the exponent as $f_{\V{s}}(p)=-\lb \Lambda-z\sqrt{R'} \rb \lb p-\hat{p}_{\V{s}} \rb
-\frac{1}{2}Q'(p-\hat{p}_{\V{s}})^2$ in \Req{X-quench} and differentiating with respect to $p$ we get the solution of the saddle-point equation $f'_{\V{s}}(p^*)=0$:
\be
p^{*}=\hat{p}_{\V{s}}-\frac{\Lambda-z\sqrt{R'}}{Q'}.
\Leq{sp-app}
\ee
As we know from the previous calculation $\frac 1N \log(\Lambda/Q')$ determines the learning edge. If $\hat{p}_{\V{s}} > \Lambda/Q'$ the dominant contribution to $p^*$ is $\hat{p}_{\V{s}}$ itself and becomes positive, implying the saddle point  is in the feasibility domain:
\be
X_{\V{s}}\approx \int_{-\infty}^{\infty} dx~ e^{f_{\V{s}}(p^{*})  -\frac{1}{2}|f''_{\V{s}}(p^{*})|x^2 }
=e^{f_{\V{s}}(p^*)}\sqrt{   \frac{2\pi}{|f''_{\V{s}}(p^*)|  }  }
=e^{ \frac{1}{2}\frac{ (\Lambda-z\sqrt{R'})^2 }{Q'} }\sqrt{   \frac{2\pi}{Q'  }  }.
\Leq{X-large}
\ee
This result coincides with \Req{X-quench} with $H(y)\to 1,\,\ y\to -\infty$. To check the validity of the saddle-point approximation we can compare the peak location $p^{*}\doteq \hat{p}_{\V{s}}$ to the width of the Gaussian, $\sigma=1 / \sqrt{|f''_{\V{s}}(p^*)|} = 1 / \sqrt{ Q' }$. Under assumption \NReq{magnitude} we  see that $\omega_{\V{s}}\leq \hat{\omega}=q'-\lambda<q'/2$ and thus $p^{*}>\sigma$ to dominant exponential--in--$N$ order. The peak height $f_{\V{s}}(p^*)$ also diverges rapidly with $N$, and the saddle-point approximation is justified. 

In the case of configurations $\V{s} \in S$ with small probabilities, $\omega_{\V{s}}>\hat{\omega}$, the dominant term to the right hand side of \Req{sp-app} is $-\Lambda/Q'$; $p^*$ becomes negative and lies outside the domain of integration. Hence, the true saddle-point value is $p^*=0$. We expand $f_{\V{s}}(p)$ around $p=0$ up to the first order, and approximate the integral as
\be
X_{\V{s}}\approx 
\int_{0}^{\infty}dp~ e^{f_{\V{s}}(0)+f'_{\V{s}}(0)p} 
=
\frac{ e^{(\Lambda-z\sqrt{R'})\hat{p}_{\V{s}}-\frac{1}{2}Q'\hat{p}_{\V{s}}^2 } }{
\Lambda-Q'\hat{p}_{\V{s}}-z\sqrt{R'}
}.
\Leq{X-small}
\ee
This is again identical to \Req{X-quench} in the limit $H(y)\approx e^{-y^2/2}/(\sqrt{2p}y)$ with $y=y_{\V{s}}-z\sqrt{R'/Q'}$. 

In addition, the saddle-point calculations above give us the expressions for the marginal measure $\rho_{\V{s}}$ given in \Rsec{learning edge}. Indeed, the marginal measure of the probability of the configuration $\V{s}$ reads 
\be
\rho_{\V{s}}(p_{\V{s}})=\frac{1}{X_{\V{s}}}\; e^{f_{\V{s}}(p_{\V{s}})} .
\ee
Therefore $\rho_{\V{s}}$ is approximately an exponentially-decaying function \NReq{dist-small} for small target probabilities,  and a Gaussian centered close to the corresponding target probability \NReq{dist-large} for large target probabilities. 

%%%%%%%%%%%%%%%%%%%%%%%%%%%%%%%%%%%%%%%%%
\subsubsection{Equations of state and their solutions}
\Lsec{EOS-quench}
Here we derive the equations of state (EOSs). Taking the derivatives of \Req{F-quench2} with respect to the order parameters, we get 
\be
&&
R'=\frac{MR}{(E+Q-R)^2},
\Leq{Rhat-quench} \ 
Q'=\frac{M}{E+Q-R},
\\
&&
Q=\lb1-2\frac{Q'}{\Lambda^2} \rb \sum_{ \V{s} \in S}\hat{p}_{\V{s}}^2-\frac{2}{\Lambda}\sum_{t\in S}\hat{p}_{\V{s}}
+e^{N \hat{\sigma} }\lb \frac{\Lambda^2+R'+Q'}{Q'^2}\rb,
\Leq{Q-quench}
\\
&&
D=Q-R=\frac{2^N}{\Lambda^2}+\frac{e^{N \hat{\sigma} }}{Q'},
\Leq{D-quench}
\\
&&
e^{N\hat{\sigma}}\frac{\Lambda}{Q'}-\frac{2^N}{\Lambda}
+\lb1-\frac{ Q' }{\Lambda^2} \rb \sum_{\V{s} \in S}\hat{p}_{\V{s}}-\frac{Q'^2}{\Lambda^3}\sum_{ \V{s} \in S}\hat{p}_{\V{s}}^2=0.
\Leq{Lambda-quench}
\ee
The order parameter $D$ defined in \Rsec{overview} naturally appears. According to assumption \NReq{magnitude}, we know that all the terms with the factor $Q'/\Lambda^2$ are not dominant in \Reqss{Rhat-quench}{Lambda-quench}; the same statement applies to the terms with the factor $e^{N \hat{\sigma} }$ in \Reqss{Q-quench}{Lambda-quench}. Matching the dominant contributions to the remaining terms we obtain
\be
&&
Q\doteq \sum_{ \V{s} \in S}\hat{p}_{\V{s}}^2\ ,
\quad  \frac{2^N}{\Lambda}\doteq  \sum_{\V{s} \in S}\hat{p}_{\V{s}}\ ,
\Leq{Q-quench-leading}
\\
&&
D\doteq \frac{2^N}{\Lambda^2} \ .
\Leq{D-quench-leading}
\ee
The physical meaning of the first equation is clear. The order parameter $Q= \sum_{\V{s}}\Ave{\lb p_{\V{s}}-\hat{p}_{\V{s}} \rb^2 }$ quantifies the average squared distance of a distribution $\V{p}$ (chosen with measure $\rho$) to the target distribution. The contributions coming from the large probabilities are negligible because, as we have seen in \Rsec{saddle-point approx}, the marginal measures for large probabilities are centered close to the target values, which makes $\sum_{\V{s} \in L} \Ave{ \lb p_{\V{s}}-\hat{p}_{\V{s}} \rb^2 }$ very small. However, for the configurations for small target probabilities, $\Ave{p_{\V{s}}^2}$ is much smaller than $\hat{p}_{\V{s}}^2$. Thus $Q$ is, to the leading order in $N$, equal to $\displaystyle{\sum_{\V{s}\in S}\hat{p}_{\V{s}}^2}$. We see, in addition, from \Req{Q-quench-leading} that $\Lambda$ ensures the normalization condition, and is equal to $2^N$ as long as most configurations are in the set $S$. According to \Req{Q-quench-leading}, we find
\be
q=r=
\left\{
\begin{array}{cc}
-\ell_2  & (\hat{\omega}<\omega_{2})   \\
\hat{\sigma}-2\hat{\omega}  &  (\omega_{2}\leq \hat{\omega} )  
\end{array}
\right.,
\Leq{q-quench}
\ \lambda=
\left\{
\begin{array}{cc}
\log 2  & (\hat{\omega}<\omega_{1})   \\
\log 2 -\hat{\sigma}+\hat{\omega}  &  (\omega_{1}\leq \hat{\omega})   
\end{array}
\right.,
\Leq{lambda-quench}
\ d=\log 2-2\lambda.
\Leq{d-quench}
\ee
The other order parameters $q',r'$ and $\hat{\omega}$ are computed accordingly. 

%%%%%%%%%%%%%%%%%%%%%%%%%%
\paragraph{Small--$E$ case.}
We can ignore $E$ when it is smaller than $D=Q-R$. In this case we get the EOSs for the parameters $q'$ and $r'$:
\be
r'=\alpha+r-2d \ ,
\Leq{rhat-quench}
\quad q'=\alpha-d,
\Leq{qhat-quench}
\ee
and the learning edge is self-consistently determined by \Req{learning edge}. Solving theses equations, we obtain
%%%%%%%%%%%%%%%%%%%%%%%%%%
\begin{description}
\item[Phase I ($\hat{\omega}<\omega_2$):]
{
\be
\lambda=\log 2, \,\, \hat{\omega}=\alpha,\,\, q'=\alpha+\log 2, \,\,q=r=-\ell_2,
\,\, r'=\alpha+2\log 2-\ell_2,\,\, d=-\log 2. 
\Leq{solution-I-quench}
\ee
}
\item[Phase II ($\omega_2 \leq \hat{\omega}<\omega_1$):]
{
\be
&&
\lambda=\log 2, \,\, \hat{\omega}=\alpha,\,\, q'=\alpha+\log 2, \,\,q=r=\sigma(\alpha)-2\alpha,
\no \\
&&
r'=\sigma(\alpha)-\alpha+2\log 2,\,\, d=-\log 2. 
\Leq{solution-II-quench}
\ee
}
\item[Phase III ($\omega_1 < \hat{\omega}$):]
{
\be
&&\lambda=\log 2+\hat{\omega}-\alpha, \,\, \hat{\omega}=\sigma^{-1}(\alpha),\,\, q'=\log 2+2\hat{\omega}-\alpha, \,\,q=r=\sigma(\hat{\omega})-2\hat{\omega},
\no \\
&& r'=2(\log 2+\hat{\omega}-\alpha)=2\lambda,\,\, d=-\log 2+2\hat{\sigma}-2\hat{\omega}. 
\Leq{solution-III-quench}
\ee
}
\end{description}
%%%%%%%%%%%%%%%%%%%%%%%%%%
The critical ratio separating phases I and II is given by $\alpha_{2}=\omega_{2}$, while the critical ratio associated to the transition from II to III is $\alpha_{1}=\omega_{1}$. We can easily check that those solutions satisfy the assumptions \NReq{magnitude}, and are therefore self-consistent.

%%%%%%%%%%%%%%%%%%%%%%%%%%
\paragraph{Large--$E$ case.}
If $E\doteq e^{N\epsilon}$ is larger than $D\doteq e^{Nd}$, {\em i.e.} if $\epsilon>d$, the EOSs of the parameters $q'$ and $r'$ are modified as
\be
r'=\alpha+r-2\epsilon \ , \quad
\Leq{rhat-quench-epsilon}
q'=\alpha-\epsilon \ .
\Leq{qhat-quench-epsilon}
\ee
These large-tolerance EOSs become valid beyond the critical values $\hat{\epsilon}(\alpha)$:
\be
\hat{\epsilon}(\alpha)=- \log 2\ {\rm (Phase \,\, I,II)}
,\,\,
\hat{\epsilon}(\alpha)=-\log 2+2\hat{\sigma}(\alpha)-2\hat{\omega}(\alpha)\ {\rm (Phase \,\, III)}.
\ee
To distinguish the solutions above from the ones of the small--$E$ case we denote phase I with large tolerance by ${\rm I}_{{\rm LT}}$ as explained in \Rsec{les3}. Phases ${\rm II}_{{\rm LT}}$ and ${\rm III}_{{\rm LT}}$ are defined in the same way. The solutions of the large-tolerance case become
\begin{description}
\item[Phase ${\rm I}_{{\rm LT}}$ ($\hat{\omega}<\omega_2$):]
{
\be
&&
\lambda=\log 2, \,\, 
\hat{\omega}=\alpha-\epsilon-\log 2,\,\, 
q'=\alpha-\epsilon=\hat{\omega}+\log 2, \,\,
q=r=-\ell_2,
\no \\
&&
\,\, 
r'=\alpha-2\epsilon-\ell_2=2\log 2+2\hat{\omega}-\ell_2-\alpha,\,\, 
d=-\log 2. 
\Leq{solution-I-quench-epsilon}
\ee
}
\item[Phase ${\rm II}_{{\rm LT}}$ ($\omega_2 \leq \hat{\omega}<\omega_1$):]
{
\be
&&
\lambda=\log 2, \,\, 
\hat{\omega}=\alpha-\epsilon-\log 2,\,\, 
q'=\alpha-\epsilon=\hat{\omega}+\log 2, 
\,\,q=r=\hat{\sigma}-2\hat{\omega},
\no \\
&&
r'=\alpha+\hat{\sigma}-2\hat{\omega}-2\epsilon
=\hat{\sigma}+2\log 2-\alpha
,\,\, 
d=-\log 2. 
\Leq{solution-II-quench-epsilon}
\ee
}
\item[Phase ${\rm III}_{{\rm LT}}$ ($\omega_1 \leq \hat{\omega}$):]
{
\be
&&\lambda=\log 2+\hat{\omega}-\hat{\sigma}, \,\, 
\hat{\sigma}-2\hat{\omega}=\log 2+\epsilon-\alpha,\,\, 
q'=\alpha-\epsilon, \,\,
q=r=\hat{\sigma}-2\hat{\omega},
\no \\
&& 
r'=\alpha+\hat{\sigma}-2\hat{\omega}-2\epsilon
=2\log 2-\alpha -\hat{\sigma}+2\hat{\omega}
,\,\, 
d=-\log 2 +2\hat{\sigma}-2\hat{\omega}. 
\Leq{solution-III-quench-epsilon}
\ee
}
\end{description}
It is instructive to examine the validity of the condition $\epsilon>d$ in phase ${\rm III}_{{\rm LT}}$. To do so we need to know the rate of increase of $\hat{\sigma}-\hat{\omega}$ with $\epsilon$ at fixed $\alpha$. Differentiating the equation determining  the learning edge in eqs.~\NReq{solution-III-quench-epsilon}, we get 
\be
\Part{( \hat{\sigma}-\hat{\omega} )}{\epsilon}{}=1+\Part{\hat{\omega}}{\epsilon}{}.
\ee
This equation can be written as 
\be
-1 \leq \Part{\hat{\omega}}{\epsilon}{}=\lb\Part{\hat{\sigma}}{\hat{\omega}}{} -2\rb^{-1}\leq -\frac{1}{2}.
\ee
The inequalities come from the condition $\partial \hat{\sigma}/\partial \hat{\omega}<1$, which holds in phase III. This implies that $d$ increases more slowly than $\epsilon$ itself in ${\rm III}_{{\rm LT}}$, since $\partial d/\partial \epsilon=2\lb 1+\partial \hat{\omega}/\partial \epsilon \rb \leq 1$. Thus the necessary condition $\epsilon>d$ is satisfied in phase ${\rm III}_{{\rm LT}}$, as it should. 
This condition will be useful for the study of the stability of the RS ansatz.

%%%%%%%%%%%%%%%%%%%%%%%%%%%%%%%%%%%%%%%%%
\subsection{Large--$\Gamma$ case}\Lsec{ME}

%%%%%%%%%%%%%%%%%%%%%%%%%%%%%%%%%%%%%%%%%
\subsubsection{Saddle-point approximation}
\Lsec{saddle-ME}
We use the saddle-point approximation in the integral defining $X_{\V{s}}$ \NReq{X-general} as in \Rsec{saddle-point approx}. We call  $f_{\V{s}}(p)$ the exponent in the integral, and write down the derivatives 
\be
&&f_{\V{s}}(p)=-\frac{1}{2}Q'(p-\hat{p}_{\V{s}})^2-(\Lambda-z\sqrt{R'})(p-\hat{p}_{\V{s}})-\Gamma p \log p,
\\
&&f'_{\V{s}}(p)=-Q'(p-\hat{p}_{\V{s}})-(\Lambda-z\sqrt{R'})-\Gamma ( \log p+1 ),
\Leq{f'-quench-MEP}
\\
&&f''_{\V{s}}(p)=-Q'-\frac{\Gamma}{p}.
\ee
The main difference with the $\Gamma=0$ case is the presence of the term $p\log p$,  singular at $p=0$. As a result the functional behavior around $p=0$  changes completely, and there is always a peak in $[0,\infty]$. This implies that the saddle-point approximation can be applied, irrespective of the target probability value. The saddle-point equation $f_{\V{s}}(p^*)=0$ can be written as
\be
p^{*}=\hat{p}_{\V{s}}-\frac{\Lambda-z\sqrt{R'}+\Gamma(1+\log p^*)}{Q'}.
\Leq{sp-quench-MEP}
\ee
Let us make the following assumption, correct when $\Gamma$ is large enough,
\be
\frac{\Lambda}{\Gamma} =O(N) .
\Leq{Gamma-largeness}
\ee
This relation, in turns, defines the meaning of `large' $\Gamma$. This assumption, combined with \Req{magnitude}, gives us  the value of the learning edge from \Req{sp-quench-MEP}: $\hat{\omega}=q'-\lambda=q'-\gamma$. For configurations with large target probabilities the dominant term is $\hat{p}_{\V{s}}$ in \Req{sp-quench-MEP}, and an iterative substitution yields
\be
p^{*}_{\V{s}} \approx \hat{p}_{\V{s}}-\frac{\Lambda-z\sqrt{R'}+\Gamma(1+\log \hat{p}_{\V{s}})}{Q'},\,\, (\V{s}\in L).
\Leq{sp-quench-MEP-large}
\ee
Expanding $f_{\V{s}}(p)$ up to the second order, we may evaluate $X_{\V{s}}$. 

For configurations corresponding to  small target probabilities, the saddle point is very different. We {\it a priori} postulate the expression of the solution 
\be
p^{*}_{\V{s}}\approx \tilde{p}+\Delta_{\V{s}},\,\, (\V{s}\in S),
\Leq{sp-quench-MEP-small}
\ee
where the dominant term $\tilde{p}$ is 
\be
\tilde{p}=e^{-1-\lb \Lambda-Q'\hat{p}_{ \V{s} }-z\sqrt{R'}  \rb/\Gamma} <\hat{p}_{\V{s}}, \,\,  (\forall\V{s} \in L) \ .
\ee
\BRsec{MEP-noconst} gives a reasoning of this form. $\tilde{p}$ is required to take this value to satisfy the normalization condition. According to the saddle-point calculation above we have
\be
1=\Ave{ \sum_{\V{s}}p_{\V{s}} }\approx \sum_{\V{s}\in S}\tilde{p}+\sum_{\V{s}\in L}\hat{p}_{\V{s}}.
\ee
For small learning edge values, $\hat{\omega}<\omega_{1}$, configurations with small probabilities dominate, implying that $\sum_{\V{s}\in S}\tilde{p} \doteq 2^{N}\tilde{p} \approx 1 \Rightarrow \tilde{p}\doteq 2^{-N}$. Thus, $\tilde{p}$ is automatically determined by the normalization condition. Moreover, the correction term $\Delta_{\V{s}}$ in \Req{sp-quench-MEP-small} is determined by expanding \Req{sp-quench-MEP} with respect to $\Delta_{\V{s}}$ to the first order and solving the resulting equation. We find
\be
\Delta_{\V{s}}=-\frac{Q'\tilde{p}^2}{Q'\tilde{p}+\Gamma}.
\Leq{Delta}
\ee
This expansion is valid if $\tilde{p}>|\Delta_{\V{s}}|$, yielding an trivial inequality $1>Q'\tilde{p}/(Q'\tilde{p}+\Gamma)$. 

%%%%%%%%%%%%%%%%%%%%%%%%%%%%%%%%%%%%%%%%%
\subsubsection{Equations of state and validity of the saddle-point approximation}

It is convenient to compute the derivatives with respect to the order parameters of the expression in \Req{F-quench}. The result is 
\be
&&Q=\sum_{\V{s}}\int Dz \Ave{\lb p-\hat{p}_{\V{s}} \rb^2}_{X_{\V{s}}},
\Leq{Q-quench-general}
\\
&&D=Q-R=\sum_{\V{s}}\int Dz \lbb \Ave{\lb p-\hat{p}_{\V{s}} \rb^2}_{X_{\V{s}}}-\Ave{ p-\hat{p}_{\V{s}} }^2_{X_{\V{s}}} \rbb,
\Leq{QR-quench-general}
\\
&&0=\sum_{\V{s}}\int Dz \Ave{p-\hat{p}_{\V{s}}}_{X_{\V{s}}}
\Leq{m-quench-general}
\ee
where we define
\be
\Ave{(\cdots)}_{X_{\V{s}}}=\frac{1}{X_{\V{s}}}\int_0^{\infty}~dp~(\cdots)e^{f_{\V{s}}(p)}.
\ee
The EOSs for $R'$ and $Q'$ are identical to \Req{Rhat-quench}, and are thus omitted above. The average value of $p$ is replaced with the saddle-point value $p^*_{\V{s}}$. We need to quantify the fluctuations around the saddle-point to estimate the terms in \Req{QR-quench-general}, that is,
\be
\Ave{\lb p-\hat{p}_{\V{s}} \rb^2}_{X_{\V{s}}}-\Ave{ p-\hat{p}_{\V{s}} }^2_{X_{\V{s}}}
\approx
\frac{ e^{f_{\V{s}}(p^*_{\V{s}})} \int dx~x^2~e^{-\frac{f''_{\V{s}}(p^*_{\V{s}})}{2}x^2  }  
}{
e^{f_{\V{s}}(p^*_{\V{s}})} \int dx~e^{-\frac{|f''_{\V{s}}(p^*_{\V{s}})|}{2}x^2}
}
=\frac{1}{|f''_{\V{s}}(p^*_{\V{s}})|}
\doteq
\lbb
\begin{array}{c}
Q'^{-1} \,\, (\V{s} \in L)
 \\
\frac{ \tilde{p} }{\Gamma}  \,\, (\V{s}\in S)
\end{array}
\right.
.
\Leq{fluctuation}
\ee
Note that, for $\V{s}\in S$, we can write $p^*_{\V{s}}-\hat{p}_{\V{s}} \doteq \tilde{p}-\hat{p}_{\V{s}}$. These considerations lead us to
\be
&&Q=\sum_{ \V{s} \in S}\int Dz (\tilde{p}-\hat{p}_{\V{s}})^2
+\sum_{ \V{s} \in L}\int Dz \lbb \lb -\frac{\Lambda+\Gamma+\Gamma \log \hat{p}_{\V{s}}}{ Q' }-z\frac{\sqrt{R'}}{Q'} \rb^2 \rbb
\no \\
&&
\doteq 
2^N \tilde{p}^2-2\tilde{p} \sum_{ \V{s} \in S}\hat{p}_{\V{s}}+\sum_{ \V{s} \in S}\hat{p}_{\V{s}}^2
+
e^{N\hat{\sigma}}\lb \frac{\Lambda+\Gamma}{Q'}\rb^2
\lbb 
\lb 1-\frac{N\hat{\omega}\Gamma}{\Lambda+\Gamma} \rb^2+\frac{R'}{(\Lambda+\Gamma)^2}
\rbb.
\Leq{Q-MEP-quench}
\\
&&
D=
\sum_{\V{s}\in S} \int Dz \frac{\tilde{p}}{\Gamma}+\sum_{\V{s}\in L}\int Dz (Q')^{-1}
\doteq
2^N\frac{\tilde{p}}{\Gamma}+\frac{e^{N\hat{\sigma}}}{Q'},
\Leq{D-MEP-quench}
\\
&&
0=
\sum_{ \V{s} \in S}\int Dz 
\lb 
\tilde{p}-\hat{p}_{\V{s}}
\rb
+
\sum_{\V{s}\in L}\int Dz 
\lb 
-\frac{\Lambda+\Gamma+\Gamma \log \hat{p}_{\V{s}}}{ Q' }-z\frac{\sqrt{R'}}{Q'}
\rb
\no \\
&&
\doteq
2^N \tilde{p}-\sum_{ \V{s} \in S}\hat{p}_{\V{s}}
-e^{N\hat{\sigma}}\lb \frac{\Lambda+\Gamma}{Q'}\rb
\lb 
1-\frac{N\hat{\omega}\Gamma}{\Lambda+\Gamma} 
\rb.
\Leq{Lambda-MEP-quench}
\ee
The dominant terms turn out to be
\be
&&
Q\doteq \sum_{\V{s}\in S}\hat{p}_{\V{s}}^2,
\Leq{Q-MEP-quench-leading}
\\
&&
D\doteq 2^N\frac{\tilde{p}}{\Gamma},
\Leq{D-MEP-quench-leading}
\\
&&
2^N\tilde{p} \doteq \sum_{\V{s}\in S}\hat{p}_{\V{s}} ,
\Leq{Lambda-MEP-quench-leading}
\ee
with the corresponding exponents,
\be
&&
q=r=
\left\{
\begin{array}{cc}
-\ell_2  & (\hat{\omega}<\omega_{2})   \\
\hat{\sigma}-2\hat{\omega}  &  (\omega_{2}\leq \hat{\omega} )  
\end{array}
\right.,
\Leq{q-MEP-quench}
\  d =
\left\{
\begin{array}{cc}
-\gamma  & (\hat{\omega}<\omega_{1})   \\
\hat{\sigma}-\hat{\omega}-\gamma &  (\omega_{1}\leq \hat{\omega})   
\end{array}
\right. ,
\Leq{d-MEP-quench}
\\
&&\lambda
= \gamma+ \frac{\log N}N + \frac 1N
\left\{
\begin{array}{cc}
\log\log 2 & (\hat{\omega}<\omega_{1})   \\
\log(\log 2+\hat{\omega}-\hat{\sigma} )&  (\omega_{1}\leq \hat{\omega})   
\end{array}
\right. .
\Leq{lambda-MEP-quench}
\ee

%%%%%%%%%%%%%%%%%%%%%%%
\paragraph{Small--$E$ case.}
Again we can neglect $E(\ll D)$. The EOSs for $q'$ and $r'$ and the equation for the learning edge are fully identical to \Req{rhat-quench} and \Req{learning edge}.  Each phase is then characterized as follows:
%%%%%%%%%%%%%%%%%%%%%%%%%%%
\begin{description}
\item[Phase ${\rm I}_{{\rm ME}}$ ($\hat{\omega}<\omega_2$):]
{
\be
\lambda
= \gamma+ \frac{\log (N\log 2)}N, \,\, \hat{\omega}=\alpha,\,\, q'=\alpha+\gamma, \,\,q=r=-\ell_2,
\,\, r'=\alpha+2\gamma-\ell_2,\,\, d=-\gamma. 
\Leq{solution-I-MEP-quench}
\ee
}
\item[Phase ${\rm II}_{{\rm ME}} $ ($\omega_2 \leq \hat{\omega}<\omega_1$):]
{
\be
&&
\lambda
= \gamma+ \frac{\log (N\log 2)}N, \,\, \hat{\omega}=\alpha, \,\, q'=\alpha+\gamma, \,\,q=r=\sigma(\alpha)-2\alpha,
\no \\
&&
r'=\sigma(\alpha)-\alpha+2\gamma,\,\, d=-\gamma.
\Leq{solution-II-MEP-quench}
\ee
}
\item[Phase ${\rm III}_{{\rm ME}}$ ($\omega_1 < \hat{\omega}$):]
{
\be
&&
\lambda
= \gamma+ \frac{\log (N(\log -\alpha+\hat \omega)2)}N, \,\, \hat{\omega}=\sigma^{-1}(\alpha),\,\, q'=\hat{\omega}+\gamma, \,\,q=r=\alpha-2\hat{\omega},
\no \\
&& r'=2\gamma,\,\, d=\hat{\sigma}-\hat{\omega}-\gamma. 
\Leq{solution-III-MEP-quench}
\ee
}
\end{description}
%%%%%%%%%%%%%%%%%%%%%%%%%%%
It is easy to check that the assumption \Req{magnitude} is satisfied by these solutions. Note that the learning edge does not depend on $\gamma$. 

We now examine the validity of the saddle-point approximation for $X_{\V{s}}$, to determine when the solutions above are correct. To do so we compare the width of the Gaussian with the location of its peak value, and check that the peak height diverges. For example, in the phase ${\rm I}_{{\rm ME}}$,
%%%%%%%%%%%%%%%%%%%%%%%%%%%
\begin{description}
\item[Peak location:]{
\be
p^*_{\V{s}}\doteq
\lbb
\begin{array}{c}
\hat{p}_{\V{s}} \doteq  e^{-N\omega_{\V{s}}}, \,\,\,\, ( \V{s} \in L)
\\
\tilde{p} \doteq e^{-N\log 2}, \,\,\,\, (\V{s} \in S)
\end{array}
\right.
.
\Leq{location-I}
\ee
}

\item[Height:]{
\be
f_{\V{s}}(p^*_{\V{s}}) \doteq 
\lbb
\begin{array}{c}
e^{N(\gamma+\hat{\omega}-2\omega_{\V{s}})}, \,\,\,\, ( \V{s} \in L)
\\
e^{N(\gamma-\log 2)}, \,\,\,\, ( \V{s} \in S)
\end{array}
\right.
.
\Leq{height-I}
\ee
}

\item[Width:]{
\be
\sigma_{\V{s}}=|f''_{ \V{s} }(  p^{*}_{\V{s}} )|^{-1/2}=\sqrt{  \frac{p^*_{\V{s}}}{  Q'p^*_{\V{s}}+\Gamma  }  }\doteq
\lbb
\begin{array}{c}
(Q')^{-1/2} \doteq  e^{  -N\frac{\gamma+\hat{\omega}}{2}  }, \,\,\,\, (\V{s} \in L)
\\
\sqrt{  \frac{ \tilde{p} }{  \Gamma  }  } \doteq e^{  -N\frac{\gamma+\log2}{2}  }, \,\,\,\, (\V{s} \in S)
\end{array}
\right.
\Leq{width-I}
.
\ee
}
\end{description}
According to these relations the saddle-point approximation is correct if $\gamma \geq \log 2\equiv \gamma_{c}$. The same approach can be applied to ${\rm II}_{{\rm ME}}$ and ${\rm III}_{{\rm ME}}$, with the resulting critical values: $\gamma_c=\log 2$ in ${\rm II}_{{\rm ME}}$ and $\gamma_c=\log 2 +\hat{\omega}-\hat{\sigma}$ for ${\rm III}_{{\rm ME}}$. These critical values are consistent with \Req{Gamma-largeness}, as $\Lambda$ is a continuous function of $\Gamma$.

This continuous change also implies that, in the $0<\Gamma<\Gamma_c=e^{N\gamma_c}$ region, all solutions coincide with the ones found for $\Gamma=0$. Consider for instance the peak value of the marginal measure over $p_{\V{s}}$ for $\V{s}\in S$. Assuming the order parameters are given by their expressions in the  $\Gamma=0$ case, we see that the condition \Req{magnitude} is satisfied, which implies that \Req{sp-quench-MEP-small} is valid. The saddle point for $\V{s}\in S$, $\tilde{p}=e^{-1- ( \Lambda-Q'\hat{p}_{ \V{s} }-z \sqrt{R'} )/ \Gamma}$, decays in a double-exponential manner with respect to $N$, as $\Lambda$ is exponentially larger than $\Gamma$. Similarly, the peak height $f_{\V{s}}(p^*_{\V{s}})$ rapidly goes to zero for $\V{s}\in S$. The peak location, $\tilde{p}$, decreases faster than the width $\sigma_{\V{s}}\doteq \sqrt{\tilde{p}/\Gamma}$. As a consequence the functional form $e^{f_{\V{s}}(p)}$ rapidly converges to a pure exponential distribution, as in the $\Gamma=0$ case.

%%%%%%%%%%%%%%%%%%%%%%%
\paragraph{Large--$E$ case.} The EOSs for $q'$ and $r'$ coincide with \Req{qhat-quench-epsilon}. The critical values of $E$ are determined by $E\doteq D$:
\be
\hat{\epsilon}(\alpha,\gamma)=-\gamma\,\, ({\rm Phase\,\, I,II}), \,\,
\hat{\epsilon}(\alpha,\gamma)=\hat{\sigma}(\alpha)-\hat{\omega}(\alpha)-\gamma\,\, ({\rm Phase\,\, III}).
\ee
New solutions can appear for $\epsilon>\hat{\epsilon}(\alpha,\gamma)$, but should be discarded as they violate the condition $E < \Gamma$, see discussion in \Rsec{role-ME}. 

%%%%%%%%%%%%%%%%%%%%%%%%%%%%%%%%%%%%%%%%%
\subsection{Stability of the replica symmetry}
\Lsec{discussion-stability}

We study the de Almeida-Thouless stability of the replica-symmetric (RS) Ansatz, see \Req{RS ansatz}, against fluctuations in the replica space \cite{ATstab}. Detailed calculations are reported in the Appendix, \Rsec{stability analysis}. The outcome of the calculation is the following stability condition against replicon fluctuations:
\be
\frac{M}{\lb E +Q-R \rb^2}
\lbb \sum_{\V{s}} \int Dz
\lb \Part{\log X_{\V{s}}}{\Lambda}{2} \rb^2
\rbb
\leq 1.
\Leq{ATcond-general}
\ee
The term in the brackets is asymptotically equivalent to $e^{N\hat{\sigma}}/Q'^2$. This can be shown in the small--$\Gamma$ case ($\Gamma \leq \Gamma_c$) using \Reqs{X-large}{X-small}, and in the large--$\Gamma$ case  ($\Gamma>\Gamma_c$) with the relation $\partial^2 \log X_{\V{s}}/ \partial \Lambda^2=\Ave{(p-\hat{p})^2}_{X_{ \V{s} }  }-\Ave{(p-\hat{p})}_{X_{ \V{s} } }^{2}  $ and \Req{fluctuation}. From \Req{Rhat-quench}, we obtain a transparent interpretation of the stability condition 
\be
\frac{M}{\lb E +Q-R \rb^2}
\lb
\frac{e^{N\hat{\sigma}}}{Q'^2} 
\rb
=
\frac{e^{N\hat{\sigma}}}{M} 
\doteq e^{N(\hat{\sigma}-\alpha)}
\leq 1.
\Leq{ATcond-simple}
\ee
Therefore the RS solution becomes unstable if the number of large-probability configurations exceeds the number of constraints. Inserting the solutions for $\hat\sigma$ in the small $E$ regime, see \Reqss{solution-I-quench}{solution-III-quench} and \Reqss{solution-I-MEP-quench}{solution-III-MEP-quench}, we find that, irrespective of the value of $\Gamma$, the RS ansatz is stable in phases I and II but is only marginally stable in phase III. Marginal stability means here that  $\hat{\sigma}$ is equal to $\alpha$ in phase III. Our calculation, limited to the leading order in $N$, cannot decide whether phase III is actually stable. 

In the large tolerance $E$ regime, the RS solution is stable across all phases, even in phase ${\rm III}_{{\rm LT}}$. Simple calculation based on \Req{solution-III-quench-epsilon} yields $\hat{\sigma}-\alpha=d-\epsilon<0$, where the last inequality is proved at the end of \Rsec{EOS-quench}.

To summarize, the RS ansatz is stable in all phases, but only marginally in phases III and ${\rm III}_{{\rm ME}}$. This result may be related to the `simple' structure of the version space. Consider the case of zero tolerance, $E=0$, and two distribution vectors, $\V{p}_{1}$ and $\V{p}_{2}$, in the version space. Any linear combination of these two vectors, $\V{p}_{t}=t\V{p}_{1}+(1-t)\V{p}_{2}$ with $t\in [0,1]$, is a normalized distribution and lies in the version space. Hence the version space is convex and connected. The instability of RS ansatz, which is usually related to the appearance of many disconnected and far apart components, may therefore not take place. 
%Meanwhile, even if the instability does not occur, the marginality of the stability condition might imply that the solution space becomes very narrow at some region inside the convex space. This may affect performance of some algorithms to solve inverse problems, although we do not have any clear indication of this at the moment. 

%%%%%%%%%%%%%%%%%%%%%%%%%%%%%%%%%%%%%%%%%
%%%%%%%%%%%%%%%%%%%%%%%%%%%%%%%%%%%%%%%%%

\section{Numerical simulations}\Lsec{numerics}

We now present a numerical procedure to sample the space of distributions with the measure $\rho$. Due to the exponential growth of the version space with $N$ this procedure is applied in practice to small values of $N\le 10$. However the results confirm the analytical calculations reported above, and provide insights about finite-size effects. We restrict to the case of zero tolerance ($E=0$) throughout this section. 

%%%%%%%%%%%%%%%%%%%%%%%%%%%%%%%%%%%%%%%%%
\subsection{Sampling algorithm}
We resort to a  Monte Carlo (MC) sampling method, in which the distribution $\V{p}$ is updated at discrete time steps. Each step corresponds to a random change of the current probability vector from $\V{p}$ to $\V{p}'=\V{p}+\Delta \V{p}$. The move $\Delta \V{p}$ must satisfy the following conditions:
\begin{description}
\item[Orthogonality to observable vectors:]{$\Delta \V{p}\cdot \V{v}^{\mu}=0, \forall{\mu}$.}
\item[Normalization:]{$\sum_{\V{s}}\Delta p_{\V{s}}=\Delta \V{p}\cdot \V{1}=0$ where $\V{1}=(1,1,\cdots,1)$.}
\item[Positivity:]{$p_{\V{s}}+\Delta p_{\V{s}}\geq 0, \forall{\V{s}}$.}
\end{description}
The orthogonality condition ensures that the constraints keep being satisfied at all steps provided they are fulfilled by the initial value of the distribution. The same statement applies to the normalization condition; we will specify below how the initial condition is chosen. The orthogonality and normalization conditions restrict the possible directions $\V{w}$ (normalized to unity) of the move. Once this direction $\V{w}$ is chosen we determine the range $[x_{min};x_{max}]$ of the allowed amplitudes $x$ of the move $\Delta \V{p}= x\; \V{w}$ to fulfill the positivity constraint: 
\be
x_{\rm min}=\min_{\V{s}}\max\lb \frac{-p_{ \V{s} }}{w_{ \V{s} } } ,\frac{1-p_{ \V{s} }}{w_{ \V{s} } } \rb
,\,\,
x_{\rm max}=\max_{\V{s}}\min\lb \frac{-p_{ \V{s} }}{w_{ \V{s} } } ,\frac{1-p_{ \V{s} }}{w_{ \V{s} } } \rb \ .
\ee
Any intermediate values between these two bounds may be chosen uniformly at random.  Next, we calculate the entropy difference $\Delta S=S(\V{p}')-S(\V{p})$ and accept the move $\V{p}\to \V{p}'$ according to the Metropolis rule, {\em i.e.} with probability
\be\Leq{pac}
p_{\rm accept}=\min\lb1,e^{\Gamma \Delta S} \rb,
\ee 
and reject it (leave $\V{p}$ unchanged) with probability $1-p_{\rm accept}$. A picture illustrating one Monte Carlo step is shown in \Rfig{algorithm}.
%%%%%%%%%%%%%%%%%%%%%%%%%%%
\begin{figure}[htbp]
\begin{center}
\includegraphics[width=0.5\columnwidth]{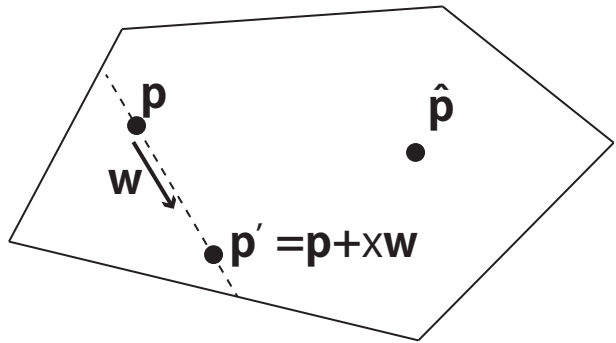}
\caption{Schematic picture of the version space and  of one Monte Carlo step. The version space is convex, and includes the target distribution $\hat{\V{p}}$. From a distribution $\V{p}$, a direction $\V{w}$ inside the solution space is randomly chosen. We choose a point on the segment along this direction (dashed line) uniformly at random, and $\V{p}$ is updated to the chosen point $\V{p}'$ with probability $p_{\rm accept}$, see \Req{pac}.  }
\Lfig{algorithm}
\end{center}
\end{figure}
%%%%%%%%%%%%%%%%%%%%%%%%%%%

To implement the algorithm, it is convenient not to  assume that the observables are Gaussianly distributed, as in the analytical treatment. Instead we consider the Fourier modes $\V{w}^{\V{k}}$ on the $N$--dimensional hypercube, where $\V{k}=(k_1,k_2,...,k_N)$, with $k_i=0,1$ for each $i$, denote the wave-number configurations. The $2^N$ components of those Fourier modes are given by
\be
\big(\V{w}^{\V{k}}\big)_{\V{s}}
=\frac{1}{ \sqrt{2^N} }\prod_{i=1}^{N}(s_{i})^{k_i} .
\ee 
The Fourier modes $\{ \V{w}^{\V{k}} \}_{\V{k}}$ form a complete orthonormal basis of the $2^N$--dimensional vector space. Note that $\V{w}^{\V{0}}=\V{1}/\sqrt{2^N}$, where $\V{0}$ is the all-zero wave-number configuration.

In the analytical calculation of \Rsec{analysis} we chose the observable $\V{v}$ to be random vectors, with Gaussian components. Any two randomly chosen vectors have very small scalar product with high probability in the large--$N$ limit. In our numerical simulation we rather choose the $M$ observables uniformly at random over the discrete set of $2^N-1$ Fourier modes $ \V{w}^{\V{k}}$, with $\V{k}\ne \V{0}$. This choice is convenient since the Fourier modes are orthogonal by construction, implying that the orthogonality condition mentioned above is automatically satisfied as soon as we choose the direction of the move $\V{w}$ as one of the Fourier modes outside the set of observables. The normalization condition is easily satisfied  as long as we exclude $\V{w}^{\V{0}}$ from the set of possible directions for the move. We are thus left with a set of $2^N-M-1$ Fourier modes, each of which is a possible direction for the Monte Carlo move. Clearly our Monte Carlo Markov Chain satisfies detailed balance and is irreducible.

%%%%%%%%%%%%%%%%%%%%%%%%%%%%%%%%%%%%%%%%%
\subsection{Results}
Using the above sampling procedure  we calculate several quantities of interest including order parameters, histograms of $p_s$, and spin-spin correlations. The target distribution we consider is again the ISM \NReq{ISM} with $H=0.5$.

%%%%%%%%%%%%%%%%%%%%%%%%%%%%%%%%%%%%%%%%%
\subsubsection{Check of equilibration}
We have run simulations for ten different samples  ($N_{\rm sample}=10$) for sizes $N=4,6$ and $8$, and for two samples ($N_{\rm sample}=2$) for size $N=10$ to compute the values of the order parameters. The error bars are estimated through the standard deviation $\sigma_{{\rm sample}}$ across the samples:
\be
{\rm Error\, bar}=\frac{\sigma_{\rm sample}}{\sqrt{N_{\rm sample}  -1} }.
\ee 
A Monte Carlo step is defined as one attempted move, irrespective of its acceptance. \Rfig{equilibration} shows the plot of the order parameter $q$ against the number of Monte Carlo steps, indicating how the system approaches equilibrium. 
%%%%%%%%%%%%%%%%%%%%%%%%%%%
\begin{figure}[htbp]
\begin{center}
\includegraphics[width=0.34\columnwidth,angle=-90]{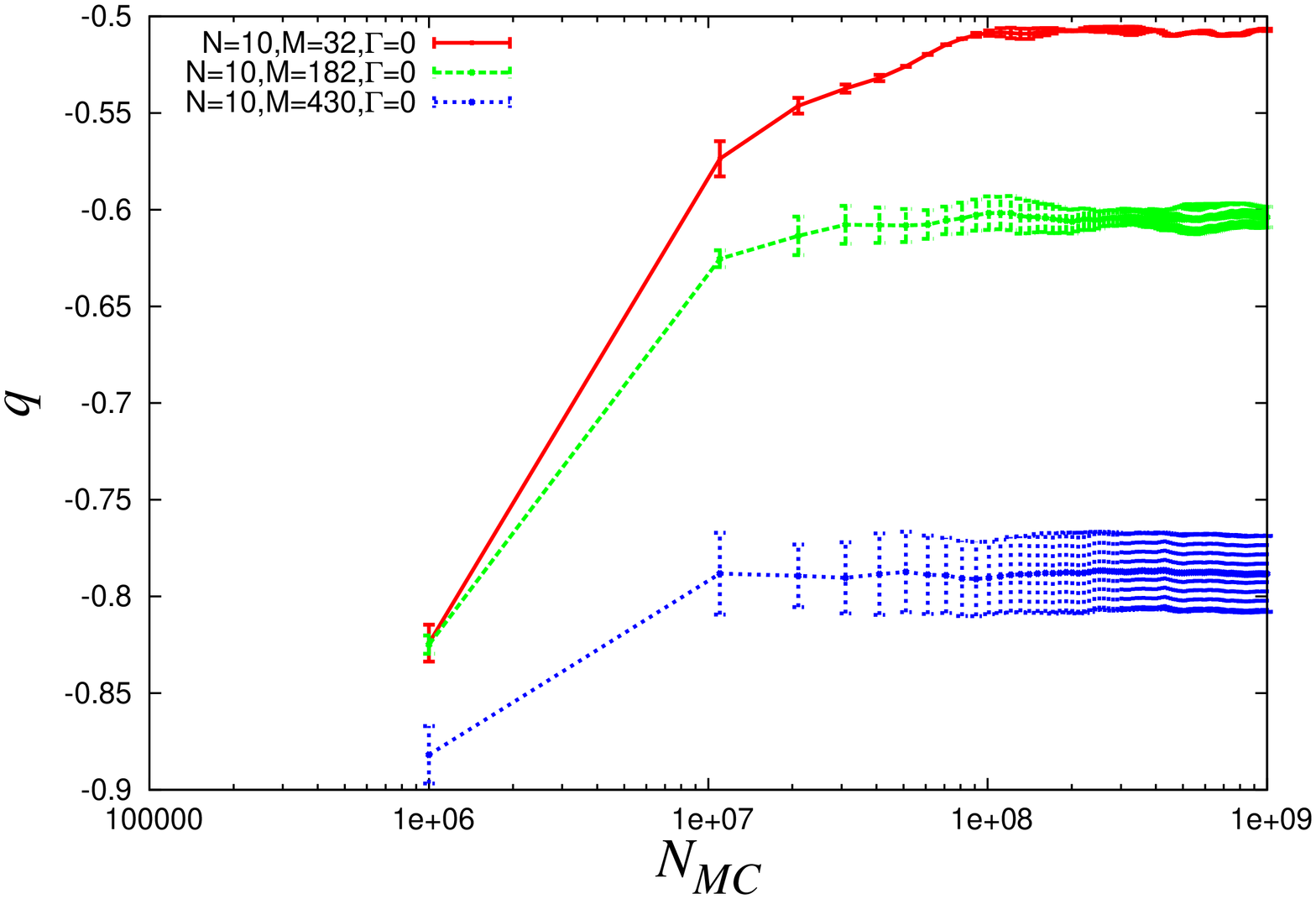}
\includegraphics[width=0.34\columnwidth,angle=-90]{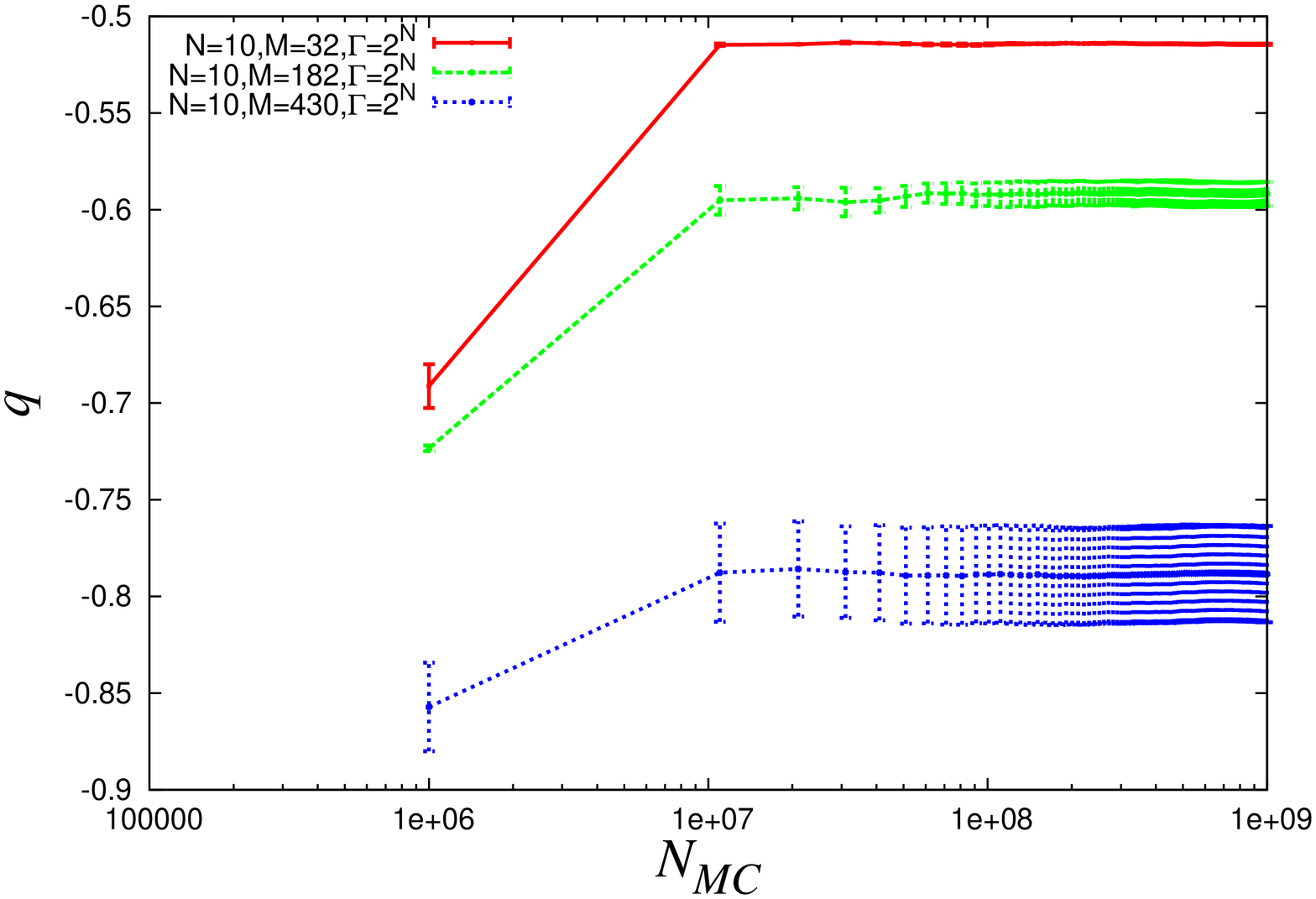}
\caption{Plots of $q$ vs. the number of Monte Carlo (MC) steps for $N=10$ spins and for different values of the number $M$ of constraints. The left panel corresponds to  $\Gamma=0$ (no entropic bias), and the right panel to $\Gamma=2^N=1024$.     }
\Lfig{equilibration}
\end{center}
\end{figure}
%%%%%%%%%%%%%%%%%%%%%%%%%%%
From the figure we see that thermalization becomes drastically faster for larger $\Gamma$. For example, simulations with $M=32$ constraints and $N=10$ spins require at least $N_{\rm MC}\approx 2\times 10^8$ steps for thermalization when $\Gamma=0$, while $N_{\rm MC}=2\times 10^7$ steps seem to be sufficient when $\Gamma \geq 2^{10}$. This trend is easy to understand, as an increase in the entropic bias concentrates more and more the measure around ME distribution, hence shrinking the space to be sampled. 

The actual  number  of Monte Carlo steps used in our simulation varied depending on the quantities we wanted to estimate. To compute the values of the order parameters we chose $N_{\rm MC}=1\times 10^7$ for $N=4$ and $6$, $N_{\rm MC}=3\times 10^7$ for $N=8$, and $N_{\rm MC}=3\times 10^8$ for $N=10$.  One third of those steps are discarded in the computation of the averages. To obtain accurate histogram of the distributions of the order parameters, however, we chose typically 5 times more MC steps.

%%%%%%%%%%%%%%%%%%%%%%%%%%%%%%%%%%%%%%%%%
\subsubsection{Order parameters, marginal measures, and spin correlations}
To confirm our analytical prediction and to estimate  finite-size effects we compute the order parameters for various values of  $M,\Gamma$ and $N$, and report the results in \Rfig{order parameters-numerics}. 
%%%%%%%%%%%%%%%%%%%%%%%%%%%
\begin{figure}[htbp]
\begin{center}
\includegraphics[width=0.4\columnwidth]{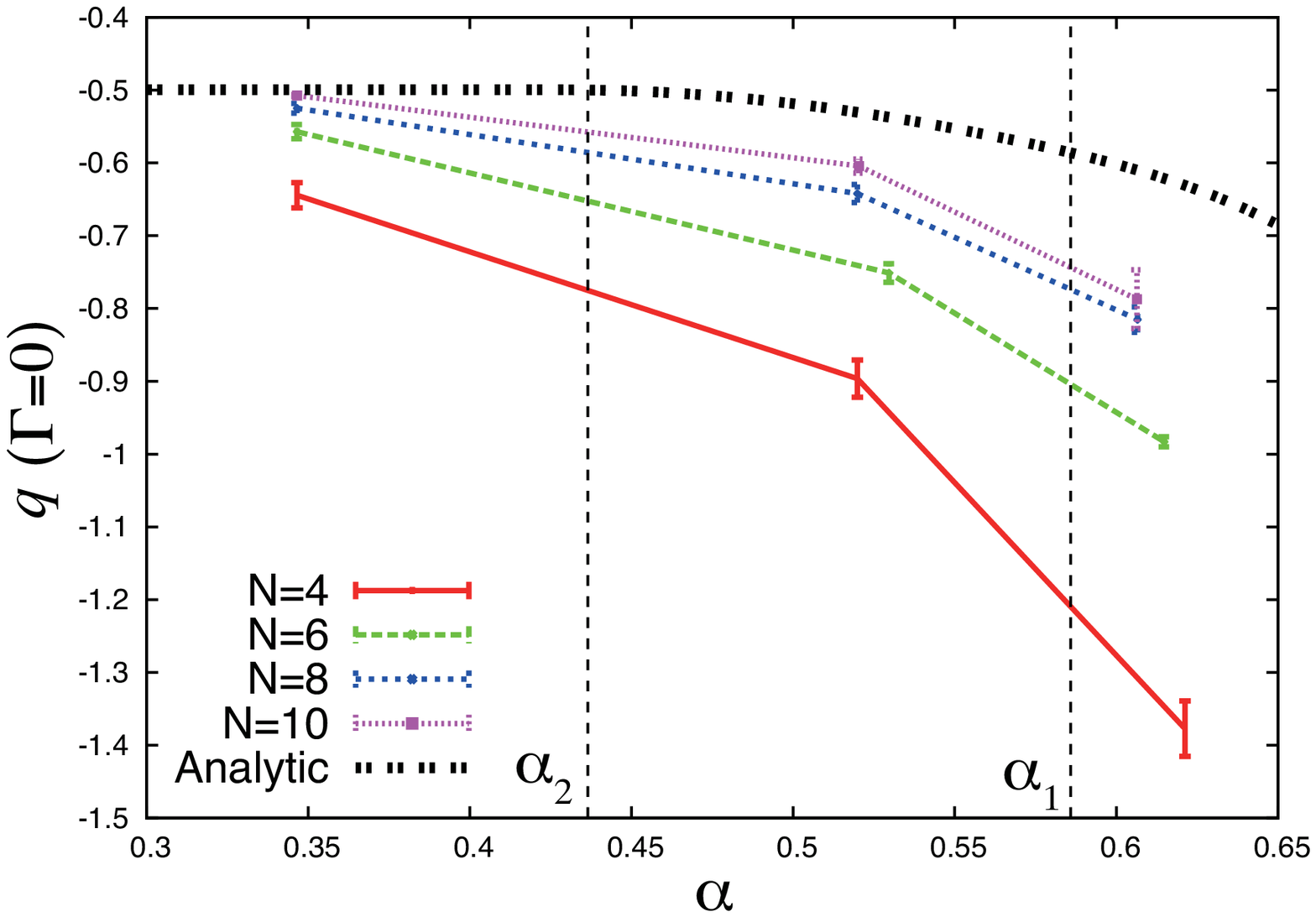}
\includegraphics[width=0.4\columnwidth]{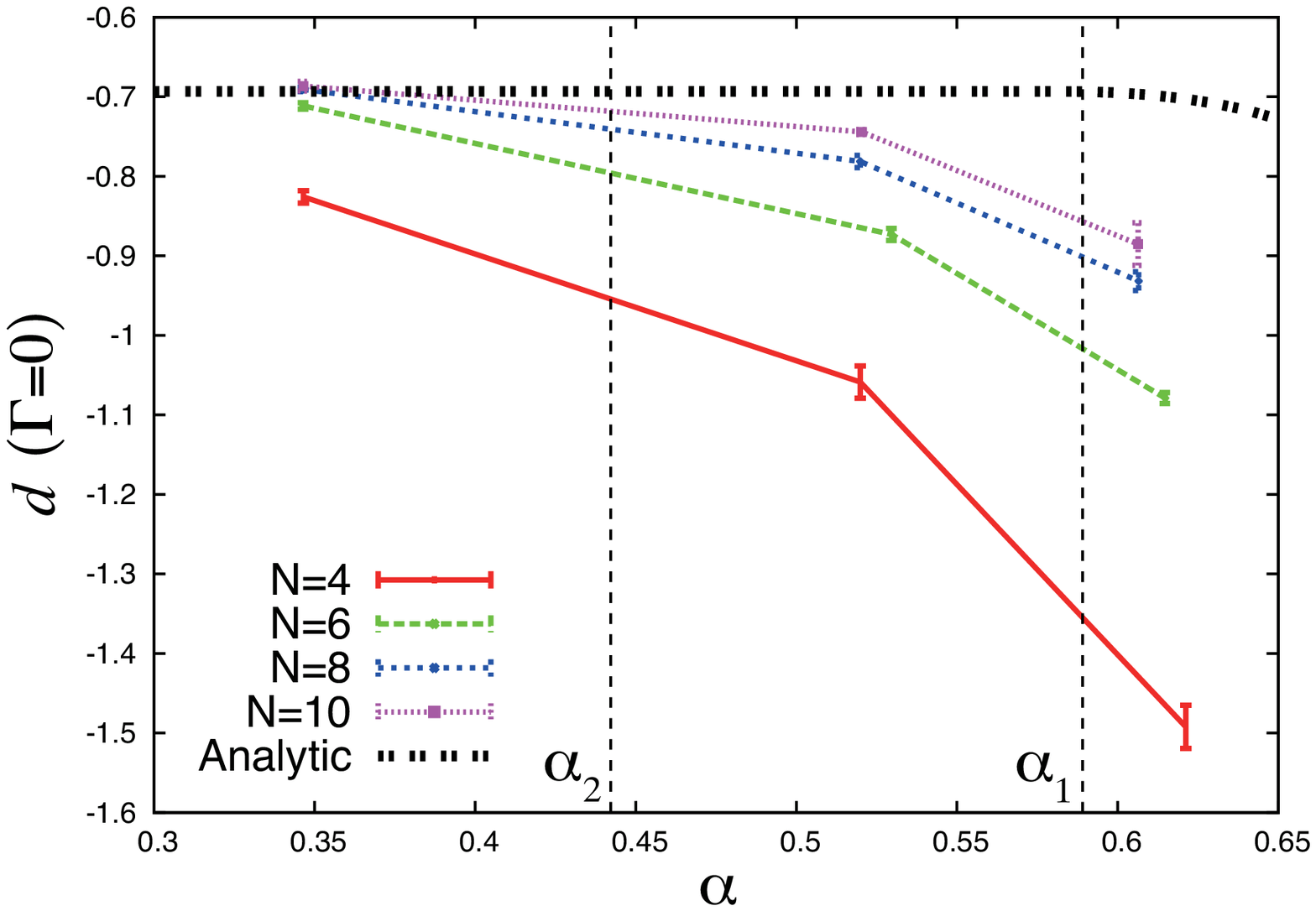}
\includegraphics[width=0.4\columnwidth]{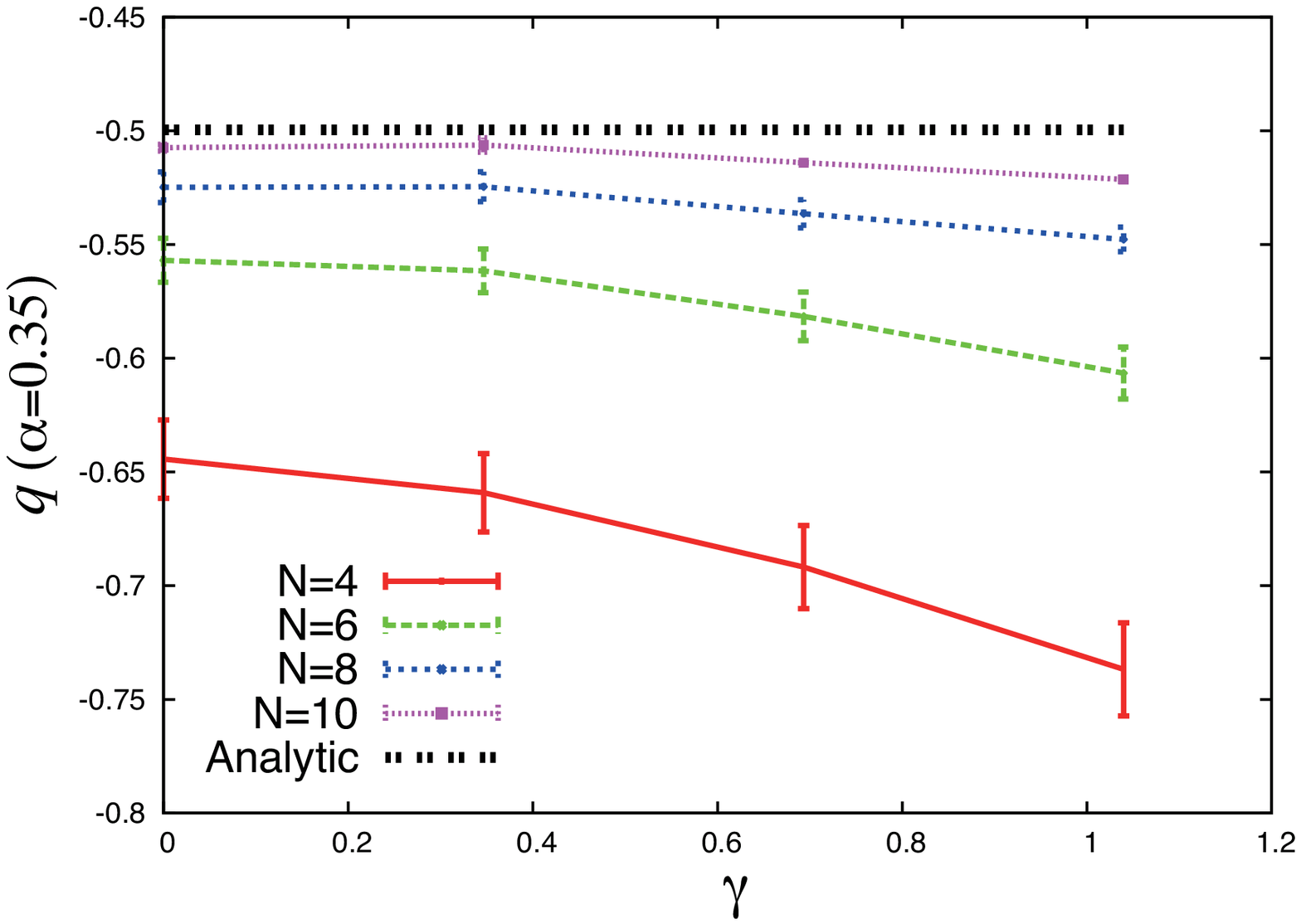}
\includegraphics[width=0.4\columnwidth]{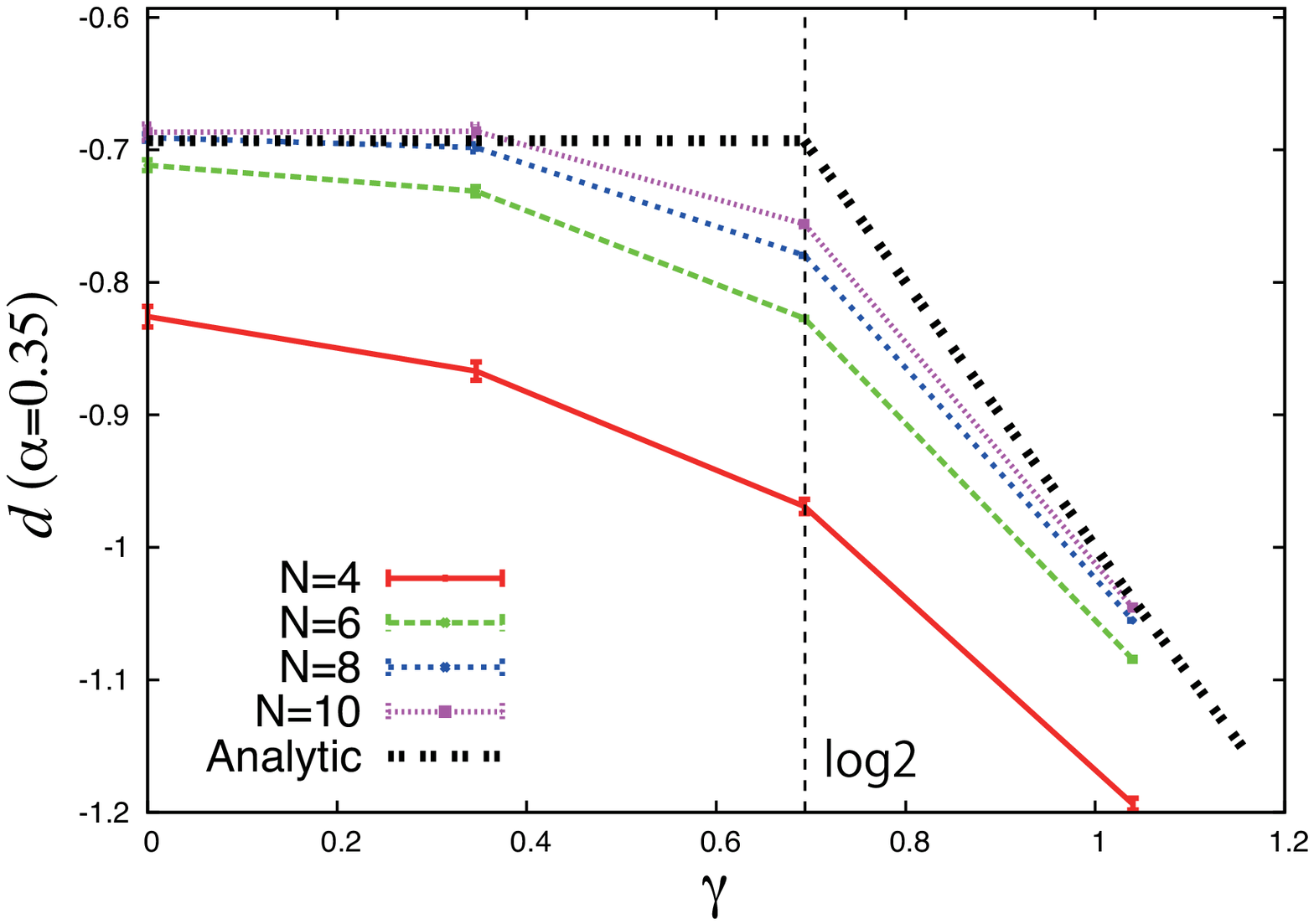}
\caption{Order parameters $q$ (left) and $d$ (right) of the ISM with $H=0.5$, computed from Monte Carlo simulations and plotted vs. $\alpha$ (upper row) and $\gamma$ (lower row). Analytical predictions are shown with the black curves. The phase transition from small to large $\Gamma$ can be best guessed in the lower, right panel. Finite-size effects seem to be stronger for larger $\alpha$. }
\Lfig{order parameters-numerics}
\end{center}
\end{figure}
%%%%%%%%%%%%%%%%%%%%%%%%%%%
Those figures clearly show that the analytical prediction, black curves, are consistent with the numerical results for sizes as small as $N=10$. The phase transitions are well captured. In particular, the phase transition to the ME phase is clearly seen in the behaviour of $d$ (right lower panel). This agreement show that our analytical findings, derived in the infinite-$N$ limit, are numerically accurate even for small sizes. 

We have also computed  the histograms of the single-configuration probabilities $p_{s}$, with the results shown in Figs.~\NRfig{pdfs} and \NRfig{pdfs-fluctuation}. Due to the symmetry of the ISM the  target probabilities $p_{\V{s}}$ depend only on the number of, say, $-1$ spins in the configuration $\V{s}$, which we call $N_s^-$. We checked that this symmetry is approximately recovered in the histograms despite the noise introduced by the Monte Carlo sampling. Therefore we show only one among the  $p_{s}$ with the same $N_{s}^{-}$ in those figures.

\Rfig{pdfs} shows the histograms corresponding to different $N_s$, for $N=8$ and $\Gamma=0$, and two values of the number $M$ of constraints. 
%%%%%%%%%%%%%%%%%%%%%%%%%%%
\begin{figure}[htbp]
\begin{center}
\includegraphics[width=0.44\columnwidth]{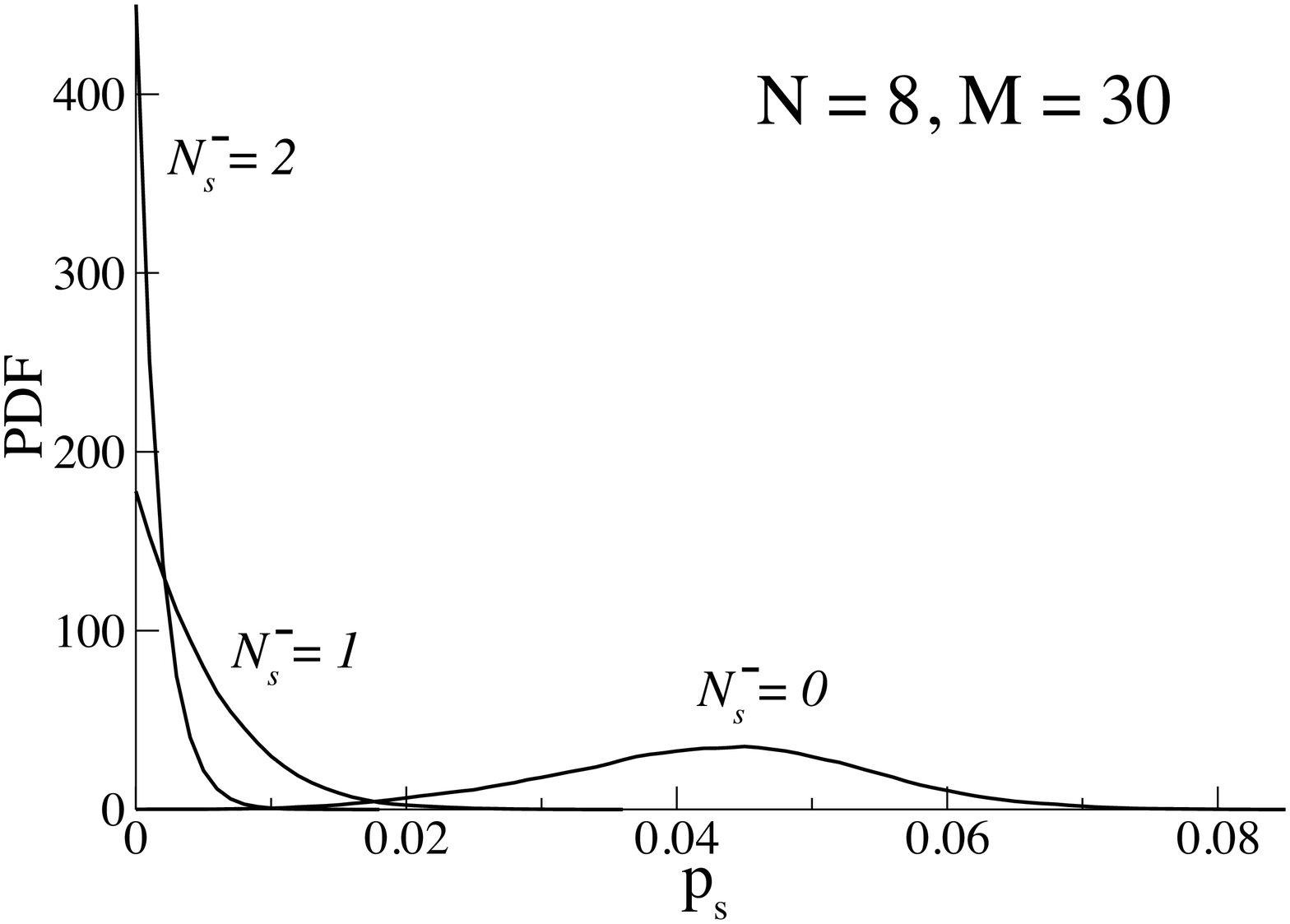}
\includegraphics[width=0.44\columnwidth]{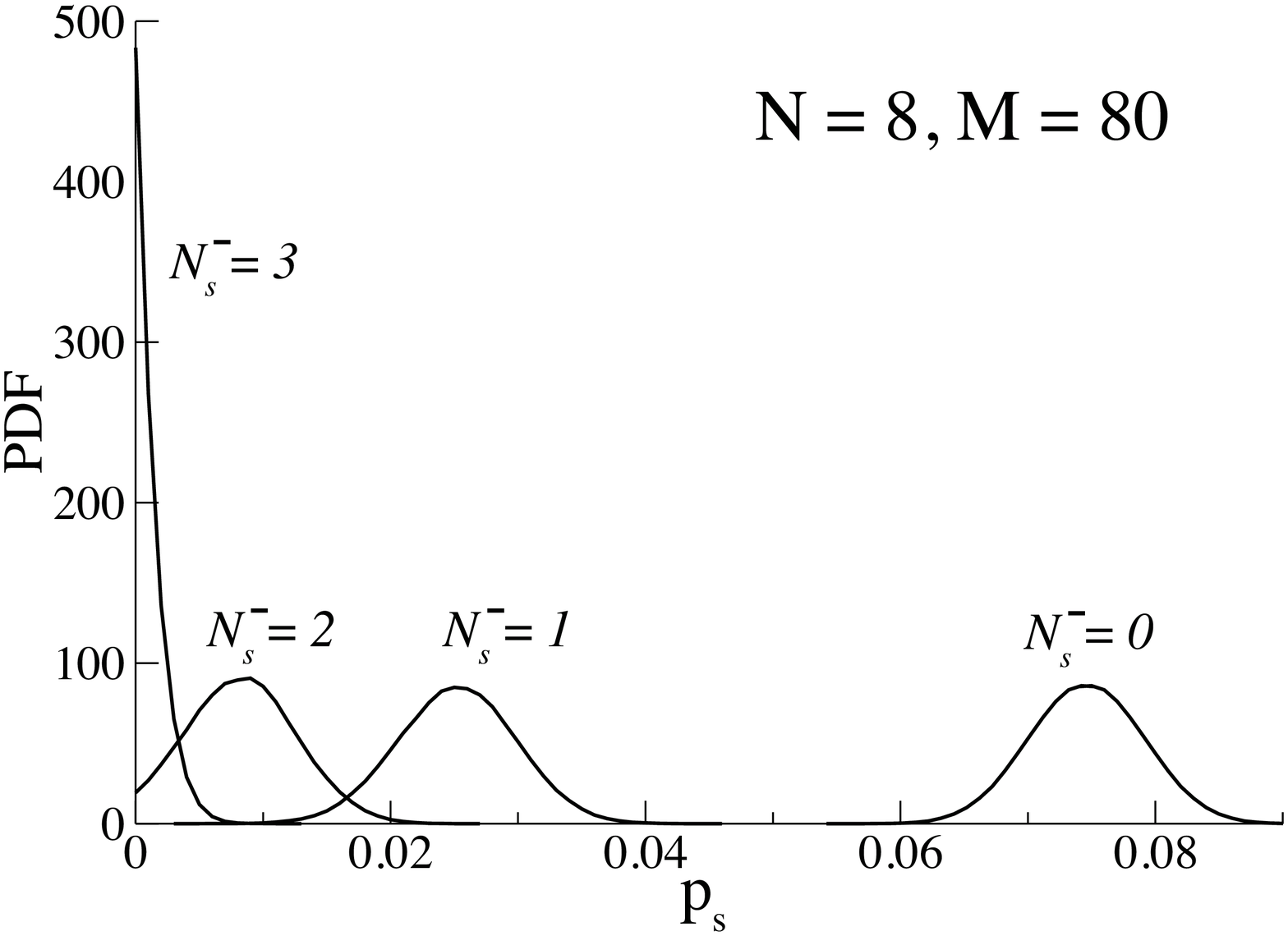}
\caption{Histograms of single-configuration probabilities $p_{\V{s}}$ for $N=8$ and $\Gamma=0$; the numbers $N_s^-$ of negative spins in the configurations are indicated on the figure. The left panel corresponds to $M=30$, and the right panel to $M=80$.}
\Lfig{pdfs}
\end{center}
\end{figure}
%%%%%%%%%%%%%%%%%%%%%%%%%%%
Two characteristic shapes of the histograms emerge. One looks like roughly Gaussian, while the other typical behaviour resembles an exponential distribution. These shapes are in very good agreement with the two possible functional dependences of $\rho_{\V{s}}(p_{\V{s}})$ upon $p_{\V{s}}$ given in \Rfig{schematic rho}. A larger number of Gaussian-like histograms is found for $M=80$ than for $M=30$, implying that more target probabilities are learned in the former case than in the latter, as expected from the theory. The size and sample dependence of the histograms of $p_0$ (corresponding to the configuration $\V{1}$ with $N_s^-=0$ minus spins) are examined in \Rfig{pdfs-fluctuation}. The corresponding target probability $\hat{p}_{\V{1}}$ is the largest one of the ISM and is  accurately learned for the values of $M$ corresponding to the figures. 
%%%%%%%%%%%%%%%%%%%%%%%%%%%
\begin{figure}[htbp]
\begin{center}
\includegraphics[width=0.44\columnwidth]{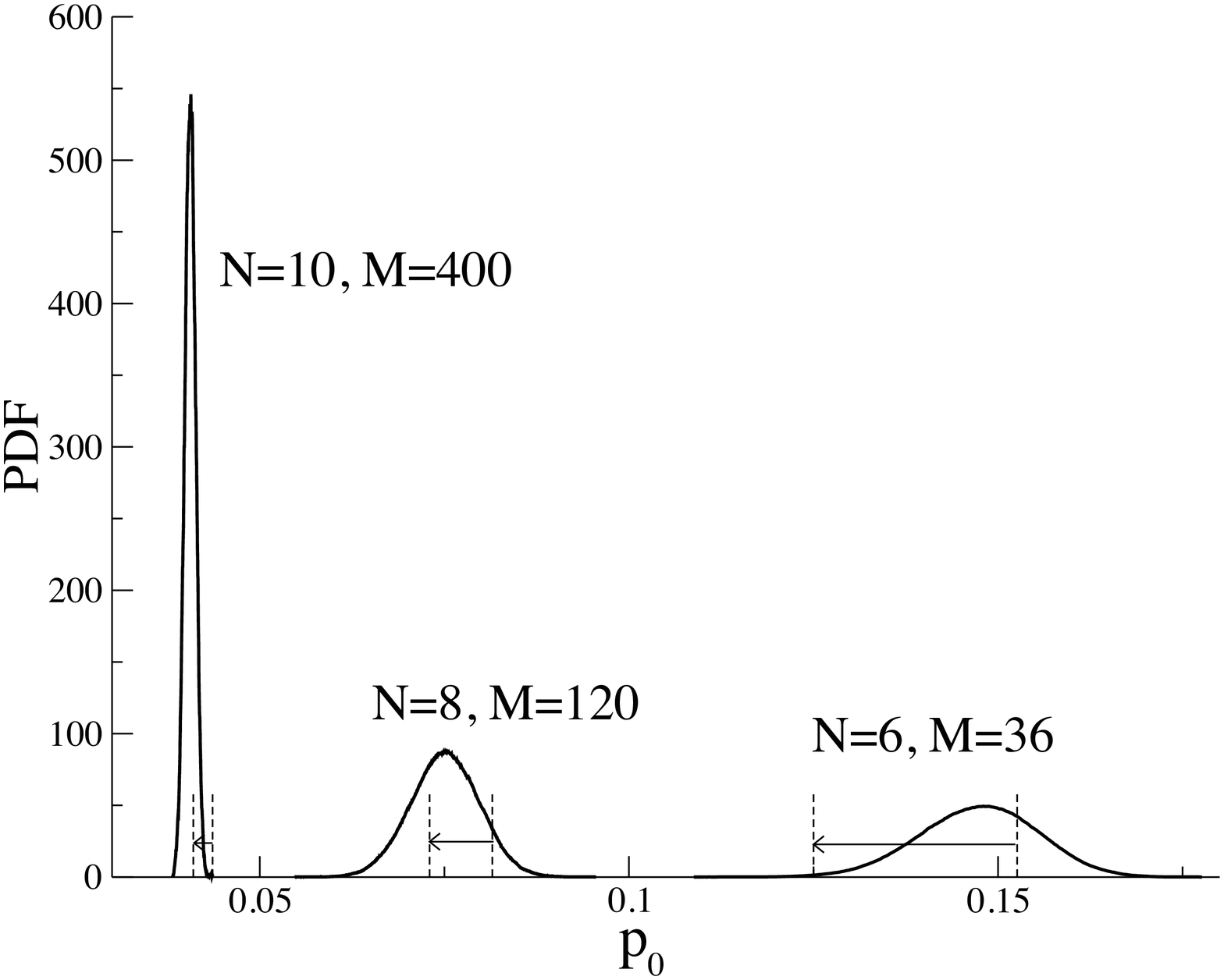}
\includegraphics[width=0.44\columnwidth]{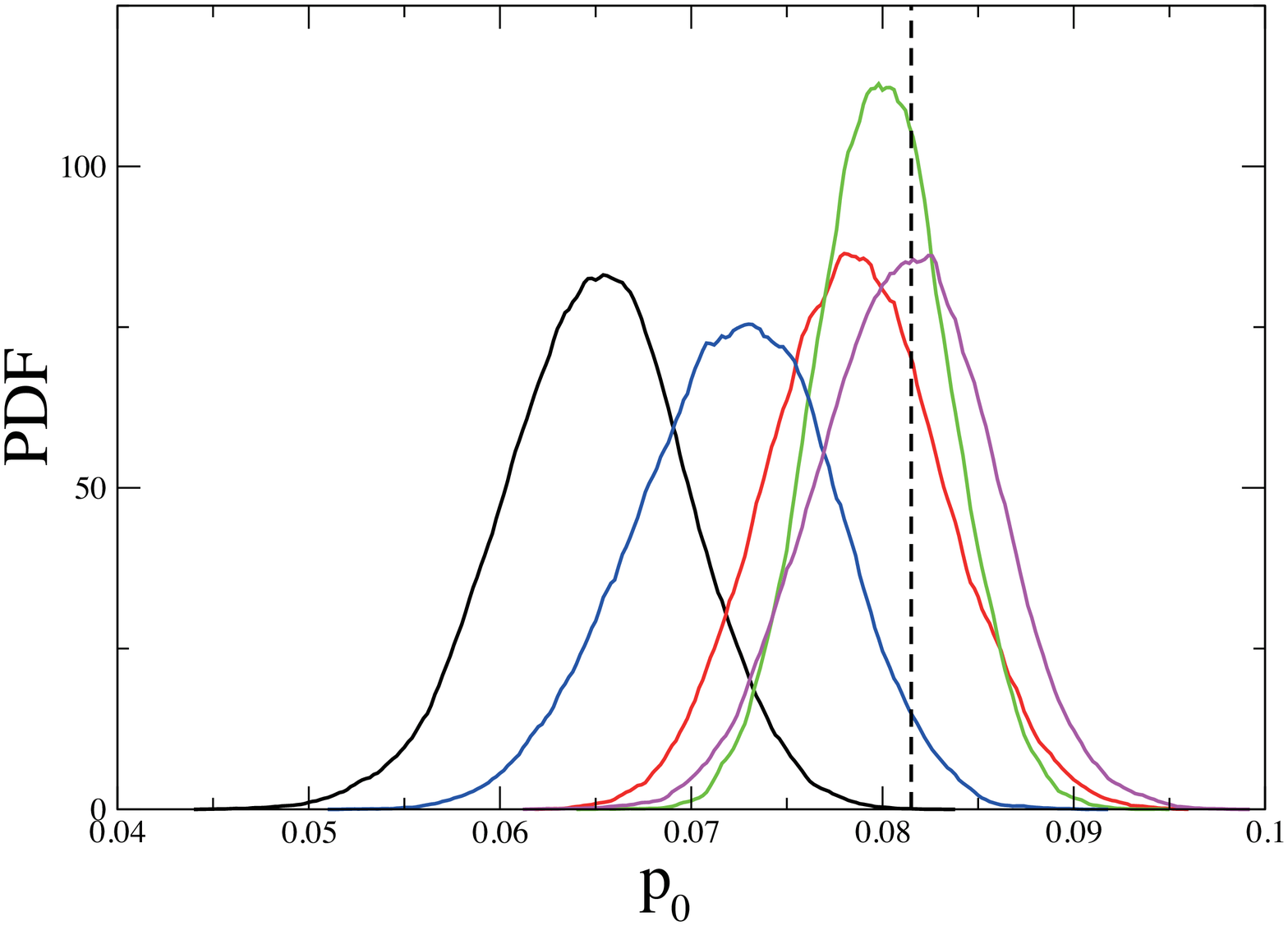}
\caption{(Left) System-size dependence of histograms of $p_0$ (corresponding to the configuration $\V{1}$ all plus spins), for fixed $\alpha\approx 0.598$ and $\Gamma=0$. For each histogram, the rightmost dashed line show the location of the target value $\hat{p}_{\V{1}}$, while the leftmost dashed line shows $\hat{p}_{\V{1}}-1/M$. (Right) Sample-to-sample fluctuations of the histogram of  $p_0$ for $N=8$, $M=80$ and $\Gamma=0$. The dashed vertical line represents the target value. }
\Lfig{pdfs-fluctuation}
\end{center}
\end{figure}
%%%%%%%%%%%%%%%%%%%%%%%%%%%
From the left panel of \Rfig{pdfs-fluctuation}, we see that our analytical prediction regarding $\rho_{\V{s}}(p_{\V{s}})$ being centered in $\hat{p}_{\V{s}}-1/M$, becomes more and more accurate as the system size increases. An excellent agreement with the the prediction is reached for $N=10$. Sample--to--sample fluctuations of the histogram of $p_{\V{1}}$ are shown in the right panel. The choice of the constraints produces moderate fluctuations in the location of the peak height of the Gaussian-like histograms. 

Last of all we show in \Rfig{pdfs-spincorrelation} the histograms of multi-spin correlations, for specific subsets of spins. We consider in particular 
\be
c_{12}=\sum_{\V{s}} p_{ \V{s} }\, s_1s_2,\,\,
c_{123}=\sum_{\V{s}} p_{ \V{s} }\, s_1s_2 s_3.
\ee
%%%%%%%%%%%%%%%%%%%%%%%%%%%
\begin{figure}[htbp]
\begin{center}
\hspace{-2mm}
\includegraphics[width=0.44\columnwidth]{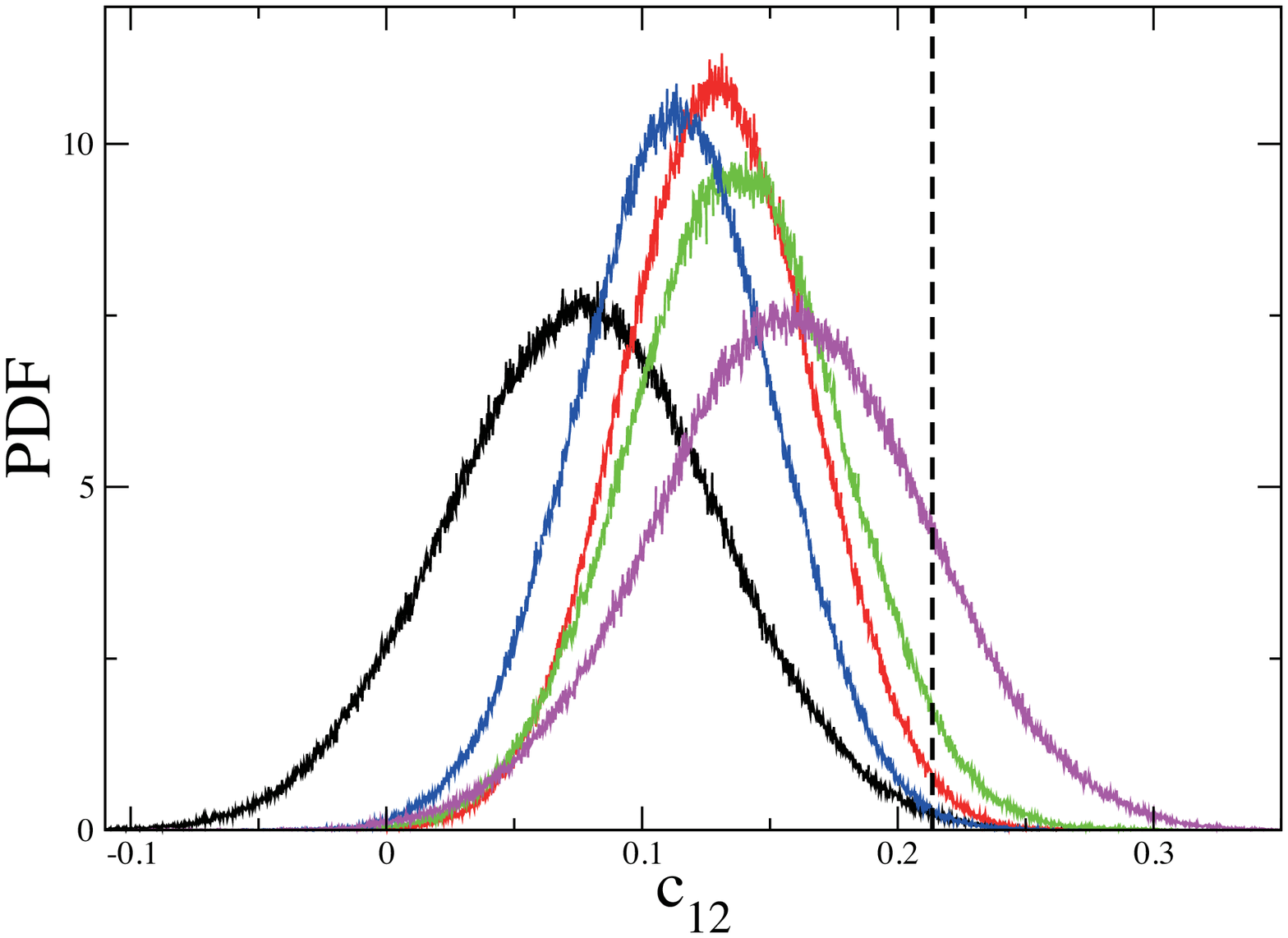}
\hspace{1mm}
\includegraphics[width=0.44\columnwidth]{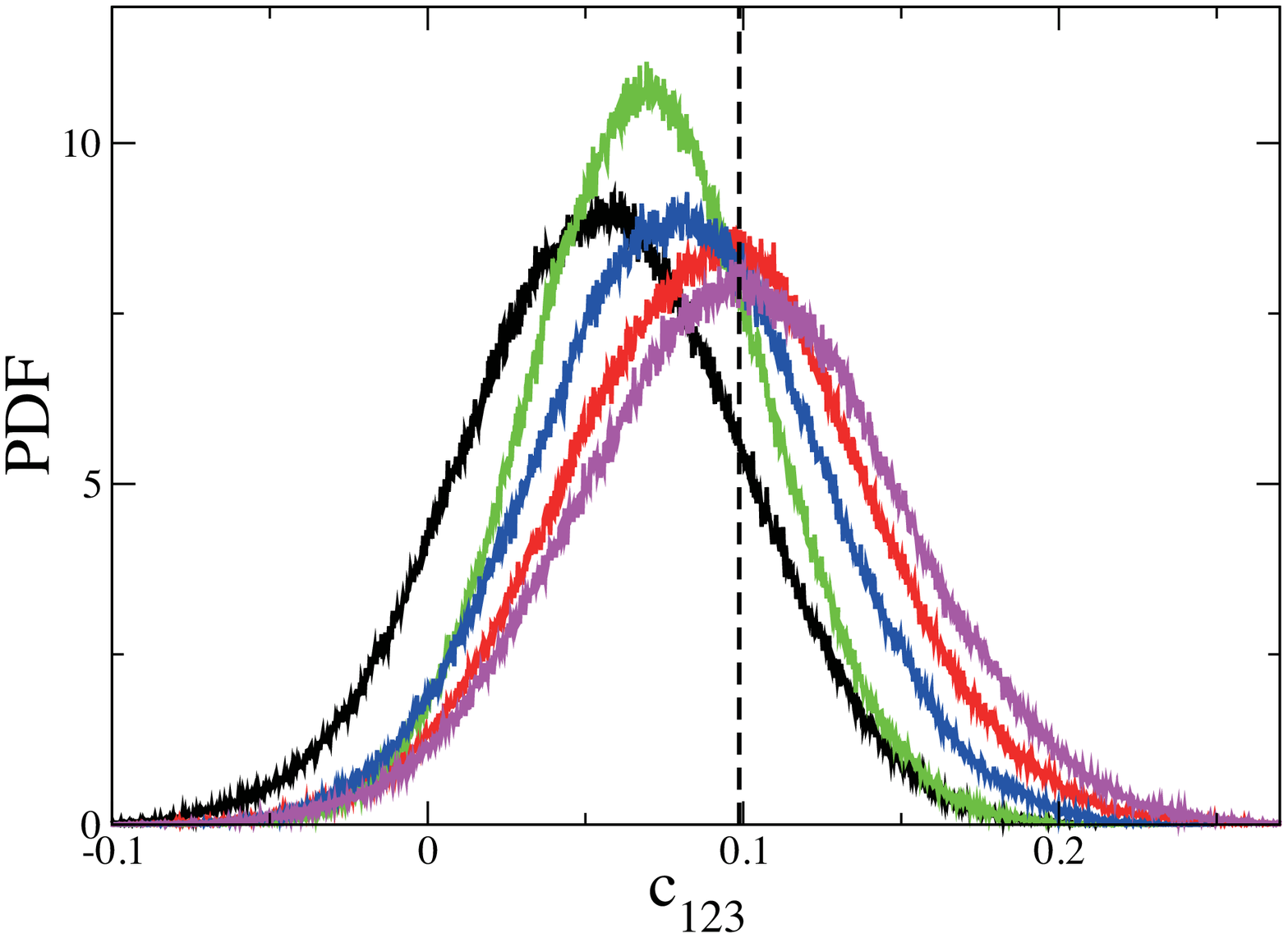}
\caption{Histograms of pairwise ($c_{12}$, left) and triplet ($c_{123}$, right) correlations for a system of $N=8$ spins. Vertical lines indicate the target distribution values. Each histogram corresponds to one sample of $M=80$ observables. }
\Lfig{pdfs-spincorrelation}
\end{center}
\end{figure}
%%%%%%%%%%%%%%%%%%%%%%%%%%%
We see that the predicted values of $c$ are in fair agreement with the target distribution values, but both the thermal and the sample-to-sample fluctuations are rather large, compared to the results of  \Rfig{pdfs-fluctuation}. 

%%%%%%%%%%%%%%%%%%%%%%%%%%%%%%%%%%%%%%%%%%%%%%%%%%%%%%%%%%%%%%%%%%%%%%%%%%%%%%%%%%
\section{Conclusion} \Lsec{concl}

In this paper we have investigated the properties of the space of probability distributions (over a large set of discrete variable configurations), constrained to reproduce many average values of observables, computed according to a target distribution $\hat{\V{p}}$.  For zero tolerance $E=0$, the space of admissible distributions $\V{p}$ defines the version space. The version space contains many distributions of interest, in addition to the target distribution $\hat{\V{p}}$, such as the center-of-mass distribution, $\langle\V{p}\rangle$, which is the flat average over all admissible distributions, and the maximum entropy distribution, $\V{p}_{ME}$. In the case of finite tolerance $E>0$ the version space acquires a probabilistic meaning. We introduce a probability measure $\rho$ over the space of distributions, giving more weight to the distributions $\V{p}$ in good agreement with the target values of the observables. The measure $\rho$ may furthermore be biased to favour distributions $\V{p}$ with large entropies $S(\V{p})$. To do so we introduce a multiplicative factor in the measure $\rho$, growing exponentially with $\Gamma\; S(\V{p})$. The coefficient $\Gamma$ acts as an inverse temperature: $\Gamma=0$ allows us to find back the unbiased measure, while, for $\Gamma\to \infty$, the measure becomes fully concentrated around $\V{p}_{ME}$. Varying the value of $\Gamma$ allows us to understand how effective is the maximum entropy principle to approximate the target distribution $\hat{\V{p}}$.

To compute analytically the volume of the version space and its main properties we have assumed that observables were quenched random variables, and we have ignored correlations between different values of the observable in different variable configurations. This assumption is not realistic, as physical observables are generally rather smooth functions of the configurations. As a result of this simplifying assumption, we have been able to compute the typical distances between the target, the centre-of-mass, and the maximum entropy distributions, as well as  the typical fluctuations around the centre of mass. The calculation was done within the replica symmetric framework, and we have checked that our solution was locally stable against replica-symmetry-breaking fluctuations (replicon modes). A major outcome of our calculation is the notion of learning edge, which separates learned from as--yet--unknown configuration probabilities. Learned probabilities correspond to large target probabilities, while the probabilities of configurations associated to small target probabilities remain unknown.  As the number of observables increases the learning edge moves to smaller and smaller probability values, implying that learning proceeds. The decrease of the learning edge is not generally accompanied by a decrease of the distance between the centre-of-mass and the target distribution, unless it hits the probabilities contributing most to that distance at specific critical points.

Numerical simulations were performed to confirm our asymptotic and analytical predictions, and to quantify  finite-size effects. Due to the exponential growth of the dimension of the version space with the number $N$ of variables we are limited to very small sizes, in practice $N\le 10$. Nevertheless results are in good agreement with our analytical results and finite-size effects do not seem to alter our asymptotic results in a significant way.  In particular our predictions for the existence of a learning edge and for the functional forms of the marginal measures over the probabilities of configurations are remarkably confirmed by the numerics. 

We have also studied the role of the entropic bias for a given number of measured observables. Our calculation shows that the maximum entropy distribution is not closer to the target distribution than any other randomly chosen distribution in the version space. This negative result is due to our choice of fully uncorrelated observables. It is quite likely that introducing an entropic bias becomes efficient and improves learning performances if we require that both the observables and the target distribution satisfy some smoothness criteria. Some preliminary results, obtained without assuming that the observable values are uncorrelated from configuration to configuration, are reported in \cite{proceedings} and seem to support this guess. Unfortunately the analytical calculations with `correlated' observables are very involved, and we have not been able to make substantial progresses so far. The theoretical importance and the practical relevance of the issues addressed here, such as how the number of constraints should depend on the smoothness of the observables and of the target distribution, or whether the phase transitions at distinct steps of the learning process found in the random uncorrelated case studied here exist in the correlated case, are strong incentives to extend this study to realistic observable ensembles. Another valuable direction for further research would be to search for `good' sets of observables. While we have restricted here to the case of $M$ independently drawn observables, it would be interesting to optimize the choice of those observables to make the volume of the version space as small as possible, and to get as close as possible to the target distribution. In this regard, we have extended the present analysis to the case of a finite number of replicas, $n\neq 0$, the result of which will be reported soon.

%%%%%%%%%%%%%%%%%%%%%%%%%%%%%%%%%%%%%%%%%%%%%%%%%%%%%%%%%%%%%%%%%%%%%%%%%%%%%%%%%%
\section*{Acknowledgments}
The authors are grateful to U. Ferrari, Y. Kabashima, J. Lebowitz, and T. Mora for fruitful discussions. T. O. acknowledges the support by Grant-in-Aid for JSPS Fellows, as well as the JSPS Core-to-Core Program ``Non-equilibrium dynamics of soft matter and information''. S. C. and R. M. acknowledge financial support from the [EU-]FP7 FET OPEN project Enlightenment 284801 and Agence Nationale de la Recherche Coevstat project (ANR-13-BS04-0012-01). Numerical calculations were partly carried out on the TSUBAME2.5 supercomputer in the Tokyo Institute of Technology.

%%%%%%%%%%%%%%%%%%%%%%%%%%%%%%%%%%%%%%%%%%%%%%%%%%%%%%%%%%%%%%%%%%%%%%%%%%%%%%%%%%
\appendix

%%%%%%%%%%%%%%%%%%%%%%%%%%%%%%%%%%%%%%%%%%%%%%%%%%%%%%%%%%%%%%%%%%%%%%%%%%%%%%%%%%
\section{Lesson from no-constraint case for finite $\Gamma$}
\Lsec{MEP-noconst}
If $\Gamma$ is finite in the no-constraint case, the volume is written as
\be
&&V(M=0,\Gamma)
=\int_{-i\infty}^{i\infty} d\Lambda~e^{\Lambda}  \int_{0}^{\infty} \prod_{\V{s}} \lb dp_{\V{s}}~e^{ -\Lambda p_{\V{s}}-\Gamma p_{\V{s}}\log p_{\V{s}} } \rb.
\Leq{volume-noconst-MEP}
\ee
Now, the saddle point with respect to $p_{\V{s}}$  is quite simple
\be
p_{\V{s}}^{*}=\tilde{p}=e^{-1-\Lambda/\Gamma}.
\ee
This is the origin of the solution assumed in \Rsec{saddle-ME}. Assuming the saddle-point approximation is correct, we can write the integration as 
\be
\int  dp_{\V{s}}~e^{ -\Lambda p_{\V{s}}-\Gamma p_{\V{s}}\log p_{\V{s}} } 
=\sqrt{2\pi}\sqrt{\frac{e^{-1-\frac{\Lambda}{\Gamma}}}{\Gamma}} e^{\Gamma e^{-1-\frac{\Lambda}{\Gamma}}}
\ee
and the log volume is
\be
F=\Lambda+2^N
\lbb
 \Gamma e^{-1-\frac{\Lambda}{\Gamma} }  
 +\frac{1}{2}
  \lb 
    \log(2\pi) -1-\frac{\Lambda}{\Gamma}-\log \Gamma
  \rb  
\rbb.
\ee
Taking a variation with respect to $\Lambda$, we get
\be
\frac{1}{2^N}=e^{-1-\frac{\Lambda}{\Gamma}}+\frac{1}{2\Gamma}.
\Leq{Lambda-noconst-MEP}
\ee
We know $\Lambda=2^N$ at $\Gamma=0$. Thus, it is natural to think that the saddle point $\tilde{p}=e^{-1-\Lambda/\Gamma}$ takes meaningful value only for $\Gamma\geq 2^N$, and otherwise $\tilde{p}$ rapidly becomes zero and the result comes back to the case $\Gamma=0$. Hence, \Req{Lambda-noconst-MEP} is satisfied in the leading scale as $2^{-N}\doteq e^{-1-\Lambda/\Gamma}$. Hence, $\Lambda=(N\log 2-1)\Gamma$ and the term $1/2\Gamma$ becomes subleading one. Substituting this, we get
\be
F \approx (N\log 2) \Gamma -2^{N}\lb N \log 2 +\log \Gamma+\frac{1}{2}(1-\log 2\pi) \rb.
\ee
The leading scale of the log volume is thus dominated by $\Gamma$. 

%%%%%%%%%%%%%%%%%%%%%%%%%%%%%%%%%%%%%%%%%%%%%%%%%%%%%%%%%%%%%%%%%%%%%%%%%%%%%%%%%%
\section{Stability analysis}
\Lsec{stability analysis}
The replica generating function is given by
\be
\phi(n)=\frac{1}{2}\sum_{a\leq b}Q_{ab}Q'_{ab}
-\frac{M}{2}\Tr{} \log  \lb E+Q\rb
+\frac{M}{2}n\log E
+\sum_{\V{s}}\log \Theta_{\V{s}},
\ee
Here $\Tr{}$ denotes the trace of matrix. We write the exponent in $\Theta_{\V{s}}$ in \Req{Theta} as 
\be
f_{\V{s}}( \{ p^a_{\V{s}}  \}_{a})=
-\sum_{a}\Lambda_a(p^{a}_{\V{s}}-\hat{p}_{\V{s}})
-\frac{1}{2}\sum_{a\leq b}Q'_{ab}(p^{a}_{\V{s}}-\hat{p}_{\V{s}})(p^{b}_{\V{s}}-\hat{p}_{\V{s}})
-\Gamma \sum_{a}p^{a}_{\V{s}}\log p^{a}_{\V{s}}.
\ee
Let us consider some small fluctuations around the RS saddle--point order parameters:
\be
Q_{ab}=Q^{{\rm RS}}_{ab}+x_{ab},\,\, Q'_{ab}=Q'^{{\rm RS}}_{ab}+2\,\hat{x}_{ab}.
\ee
Then,
\be
&&
\frac{1}{2}\sum_{a\leq b}Q_{ab}Q'_{ab}
\approx 
\frac{1}{2}\sum_{a\leq b}Q^{{\rm RS}}_{ab}Q'^{{\rm RS}}_{ab}
+
\sum_{a\leq b}x_{ab}\hat{x}_{ab}\ ,
\\
&&
\Tr{}\log \lb E+Q\rb
\approx
\Tr{}\log \lb E+Q^{{\rm RS}}\rb
-\frac{1}{2}\Tr{}(Ax)^2\ .
\ee
where we define $A=\lb E+Q^{{\rm RS}} \rb^{-1}$, and 
\be
f_{\V{s}}=
f^{{\rm RS}}_{\V{s}}-\sum_{a \leq b}\hat{x}_{ab}\bar{p}^{a}_{\V{s}}\bar{p}^{b}_{\V{s}}
=
f^{{\rm RS}}_{\V{s}}+\Delta f_{\V{s}} 
,
\ee
with $\bar{p}^{a}_{\V{s}}=p^{a}_{\V{s}}-\hat{p}_{\V{s}}.$ Thus
\be
\log \Theta_{\V{s}}
\approx
\log \Tr{}e^{f^{{\rm RS}}_{\V{s}}}\lb 1+\Delta f_{\V{s}}+\frac{1}{2}\Delta f_{\V{s}}^2\rb
\approx
\log \Theta^{{\rm RS}}
+\frac{1}{2}\Ave{   \Delta f^2_{\V{s}} }  -\frac{1}{2}\Ave{\Delta f_{\V{s}}}^{2}_{\V{s}},
\ee
where we have introduced the notation
\be
\Ave{\cdots}_{\V{s}}=\frac{1}{\Theta^{{\rm RS}}_{\V{s}}}\Tr{}e^{f^{{\rm RS}}_{\V{s}}}(\cdots).
\ee
Note that we have omitted all the first-order terms, which vanish at the saddle point. Thus we get $\phi(n)=\phi^{{\rm RS}}+\Delta$ with
\be
&&
\hspace{-10mm}
2\Delta
=2\sum_{a \leq b}x_{ab}\hat{x}_{ab}+\frac{M}{2}\Tr{}(Ax)^2
+\sum_{a \leq b}\sum_{c \leq d}\hat{x}_{ab}\hat{x}_{cd}
\sum_{\V{s}}
\lb
\Ave{\bar{p}^{a}_{\V{s}}\bar{p}^{b}_{\V{s}}\bar{p}^{c}_{\V{s}}\bar{p}^{d}_{\V{s}}}_{\V{s}}
-\Ave{\bar{p}^{a}_{\V{s}}\bar{p}^{b}_{\V{s}}}\Ave{\bar{p}^{c}_{\V{s}}\bar{p}^{d}_{\V{s}}}_{\V{s}}
\rb.
\Leq{AT-quadratic}
\ee
We write $A_{aa}\equiv X$ and $A_{ab}\equiv Y\,\, (a\neq b)$, and those components are easily calculated 
\be
&&
X=\frac{E+Q+(n-2)R
}{
\lb E+Q+(n-1)R \rb \lb E+Q-R \rb
},\hspace{2mm}
Y=-\frac{R}{\lb E+Q+(n-1)R \rb \lb E+Q-R \rb},
\\ 
&&
X+(n-1)Y=\frac{1}{\lb E+Q+(n-1)R \rb},\hspace{2mm}
X-Y=\frac{1}{\lb E+Q-R \rb}.
\ee
\BReq{AT-quadratic} is a quadratic form with respect to $ \{x_{ab}\}$ and $\{\hat{x}_{ab}\}$ and can be written as $2\Delta=\V{V}^{t}G\V{V}$, where $\V{V}$ is a column vector with components $\{x_{ab}\}$ and $\{ \hat{x}_{ab}\} $, and $\V{V}^{t}$ is its transpose. We now want to determine whether the Hessian matrix $G$ has unstable modes. The most likely candidate lies in the replicon eigenspace, which is spanned by the vectors whose components ($x_{ab},\hat x_{ab}$) depend only on whether their replica indices are equal or different to two fixed values $a=\theta$ and $b=\eta$ \cite{ATstab}. In the replicon space, diagonal ($a=b$) fluctuations tend to be irrelevant since other (transverse, longitudinal) modes  span the $n$-dimensional space ($x_{aa},\hat x_{aa}$). Hence we set hereafter $x_{aa}=\hat{x}_{aa}=0$.

We arrange the components of $\V{V}$ by ordering $x$ as
\be
\V{v}=(x_{12},x_{13},\cdots,x_{1n},x_{23},\cdots,x_{n-1,n}),
\ee
and $\hat x$ as well. 
The Hessian $G$ has the following form
\begin{equation}
G=
 \left(\begin{array}{c|c}	      
   \begin{array}{ccc}
    S & T\cdots T & U\cdots U \\
  & \ddots & \\
 T\cdots T  & U\cdots U & S
\end{array}                      & I \\ \hline
I & 	\begin{array}{ccc}
    \hat{S} & \hat{T}\cdots \hat{T} & \hat{U}\cdots \hat{U} \\
  & \ddots &  \\
 \hat{T} \cdots \hat{T}  & \hat{U}\cdots \hat{U} & \hat{S}
 \end{array}
\end{array}
 \right), \Leq{eq:G}
\end{equation}
where $I$ is the identity matrix, and 
%$S$ and $\hat{S}$ are the diagonal, $T$ and $T'$ are the semi-diagonal, $U$ and $U'$ are the non-diagonal, contributions of $\Tr{}(Ax)^2$ and $C_{ab,cd}$, respectively. $I$ is the identity matrix. Each component is 
\begin{eqnarray}
&&
\hat{S}=\sum_{\V{s}} \lb \Ave{(\bar{p}^{a}_{\V{s}}\bar{p}^{b}_{\V{s}})^2}_{\V{s}}-\Ave{\bar{p}^{a}_{\V{s}}\bar{p}^{b}_{\V{s}}}_{\V{s}}^2 \rb, \\
&&
\hat{T}=\sum_{\V{s}} \lb \Ave{(\bar{p}^{a}_{\V{s}})^2\bar{p}^{b}_{\V{s}}\bar{p}^{c}_{\V{s}}}_{\V{s}}-\Ave{ \bar{p}^{a}_{\V{s}}\bar{p}^{b}_{\V{s}} }_{\V{s}}
\Ave{ \bar{p}^{a}_{\V{s}}\bar{p}^{c}_{\V{s}} }_{\V{s}} 
\rb,\\
&&
\hat{U}=\sum_{\V{s}} \lb \Ave{\bar{p}^{a}_{\V{s}}\bar{p}^{b}_{\V{s}}\bar{p}^{c}_{\V{s}}\bar{p}^{d}_{\V{s}}}_{\V{s}}-\Ave{\bar{p}^{a}_{\V{s}}\bar{p}^{b}_{\V{s}}}_{\V{s}}\Ave{\bar{p}^{c}_{\V{s}}\bar{p}^{d}_{\V{s}}}_{\V{s}} \rb \label{eq:Rhat}, \\
&&S= M(X^2+Y^2)\ , \quad T=MY(X+Y)\ ,\quad  U=2MY^2 \ . \label{eq:P}
\end{eqnarray}
Let us find the eigenvectors of this matrix.
The first eigenvector $\V{V}_{1}$ is obtained by assuming
$x_{ab}=a$ and $\hat{x}_{ab}=\hat{a}$
for any $\lambda,\nu$.
The upper half (in the matrix drawn in \Req{eq:G}) of the eigenvalue equation 
$G\V{V}_{1}=\lambda_{1}\V{V}_{1}$
 gives
\begin{equation}
\left(
S+2(n-2)T+\frac{1}{2}(n-2)(n-3)U
\right)a+\hat{a}=\lambda_{1}a,
\end{equation}
and the lower half yields
\begin{equation}
a+\left(
\hat{S}+2(n-2)\hat{T}+\frac{1}{2}(n-2)(n-3)\hat{U}
\right)\hat{a}=\lambda_{1}\hat{a}.
\end{equation}
These equations have the eigenvalue 
$\lambda_{1}$ with non-vanishing $a,\hat{a}$, 
which must satisfies the following relation
\begin{equation}
\lambda_{1}^2-(C_{1}+\hat{C}_{1})\lambda_{1}+(C_{1}\hat{C}_{1}-1)=0\label{eq:lambda1},
\end{equation}
where $C_{1}=S+2(n-2)T+(n-2)(n-3)U/2$ and 
$\hat{C}_{1}=\hat{S}+2(n-2)\hat{T}+(n-2)(n-3)\hat{U}/2$.
This equation says that this mode spans a two-dimensional space.

The next type of solution $\V{V}_{2}$ is obtained by selecting a replica index, say, $\theta=1$.
This solution $\V{V}_{2}$ has 
$x_{ab}=b$ and $\hat{x}_{ab}=\hat{b}$
when $\lambda$ or $\nu$ is equal to $1$, 
$y^{\lambda\nu}=c$ and $\hat{y}^{\lambda\nu}=\hat{c}$ 
otherwise.
The first row of the eigenvalue equation 
$G\V{V}_{2}=\lambda_{2}\V{V}_{2}$
gives
\begin{equation}
(S+(n-2)T)b+
\left(
(n-2)T +
\frac{1}{2}(n-2)(n-3)U
\right)c+\hat{b}=\lambda_{2}b,
\end{equation}
and the first row of the lower half of matrix $G$ in \Req{eq:G} yields
\begin{equation}
b+\left(
\hat{S}+(n-2)\hat{T}
\right)\hat{b}
+
\left(
(n-2)\hat{T}
+\frac{1}{2}(n-2)(n-3)\hat{U}
\right)\hat{c}=\lambda_{2}\hat{b}.
\end{equation}
We now impose the orthogonality condition over $\V{V}_{2}$ and $\V{V}_{1}$.
%We expect that  the upper half of $\V{V}_{2}$ is orthogonal with that of $\V{s}_{1}$ independent of the lower half space.
This leads to
\begin{equation}
(n-1)b+\frac{1}{2}(n-1)(n-2)c=0,
\Leq{orthogonal-12}
\end{equation}
and the same relation holds for $\hat{b}$ and $\hat{c}$.
Substituting these conditions, we get
\begin{eqnarray}\left(
S+(n-4)T-(n-3)U
\right)b
+\hat{b}=\lambda_{2}b\ ,\  \left(
\hat{S}+(n-4)\hat{T}-(n-3)\hat{U}
\right)\hat{b}
+b=\lambda_{2}\hat{b}\ ,
\end{eqnarray}
 which leads to
\begin{equation}
\lambda_{2}^2-(C_{2}+\hat{C}_{2})\lambda_{2}+(C_{2}\hat{C}_{2}-1)=0
\label{eq:lambda2},
\end{equation}
where $C_{2}=S+(n-4)T-(n-3)U$ and 
$\hat{C}_{2}=\hat{S}+(n-4)\hat{T}-(n-3)\hat{U}$.
As there are $n$ possible choices for the replica index $\theta$
and  two eigenvalues/eigenvectors $\V{V}_{1}$
and $\V{V}_{2}$ for each choice, this particular subspace is of dimension $2n$.

The third mode $\V{V}_{3}$ is obtained by treating two replicas $\theta,\omega$ as special ones. This solution $\V{V}_{3}$ has 
$x_{\theta\omega}=d$ and $\hat{x}_{\theta\omega}=\hat{d}$,
$x_{\theta a}=e$ and $\hat{x}_{\omega a}=\hat{e}$,
and 
$x_{ab}=f$ and $\hat{x}_{ab}=\hat{f}$
otherwise. 
The orthogonality condition with $\V{V}_2$ is 
\be
b(d+(n-2)e)+c\lb(n-2)e+\frac{1}{2}(n-2)(n-5)f \rb
=
b\lb d+(n-4)e-(n-5)f \rb=0,
\ee
where we use \Req{orthogonal-12}. The one with $\V{V}_1$ is
\be
d+2(n-2)e+\frac{1}{2}(n-2)(n-5)f=0.
\ee
These relations mean
\be
e=-\frac{d}{n-2},\,\, \frac{1}{2}(n-2)(n-5)f=-d-2(n-2)e=d.
\ee
Similar relations hold for the hat variables. According to these relations, the first rows of the upper and lower halves of the eigenvalue equation $G\V{V}_3=\lambda_3\V{V}_3$ yield 
\be
&&dS+2(n-2)eT+\frac{1}{2}(n-2)(n-5)fU+\hat{d}
=(S-2T+U)d+\hat{d}=\lambda_3 d,
\\
&&d+(\hat{S}-2\hat{T}+\hat{U})\hat{d}=\lambda_3 \hat{d},
\ee
Thus we obtain
\begin{equation}
\lambda_{3}^2-(C_{3}+\hat{C}_{3})\lambda_{3}+(C_{3}\hat{C}_{3}-1)=0
\label{eq:lambda3},
\end{equation}
where $C_{3}=S-2T+U$ and 
$\hat{C}_{3}=\hat{S}-2\hat{T}+\hat{U}$.
This solution spans a $n(n-3)$-dimensional space, implying that no eigenmode are left. 

For the stability of the saddle point, 
all of the eigenvalues must be non-negative. 
This condition corresponds to 
\begin{equation}
\forall{i},\hspace{5mm}C_{i}\;\hat{C}_{i}\leq 1.\label{eq:positivity}
\end{equation}
This condition takes into account the fact that $Q'_{ab}$ is originally 
a pure imaginary variable, which means that 
$\delta Q_{ab}\delta Q'_{ab}$ is  
associated with a multiplicative factor $i$ and
$\delta Q'_{ab}\delta Q'_{ab}$
acquires a factor $-1$. Hence, if we change variable 
from $Q'_{ab}$ to $i\, Q'_{ab}$, 
the diagonal block in the lower half part of $G$ gets a factor $-1$,
and the off-diagonal part becomes $i\, I$, which leads to 
the positivity condition in eq.\ (\ref{eq:positivity}). 

The replicon mode corresponds to $\lambda_3$ and $\V{V}_3$. Thus, the stability condition is 
\be
(S-2T+U)(\hat{S}-2\hat{T}+\hat{U})\leq 1.
\ee
A straightforward calculation shows that
\be
S-2T+U=M(X-Y)^2=\frac{M}{(E+Q-R)^2}.
\ee
We now turn to the calculation of $\hat{S}-2\hat{T}+\hat{U}$. Within the RS assumption the average $\Ave{O}_{\V{s}}$, where $O=O_aO_b\cdots O_{c}$, can be expressed as
\be
\Ave{O}_{\V{s}}=
\frac{1}{\int Dz X_{\V{s}}^n}\int Dz X_{\V{s}}^n\Ave{O_a}_{X_{\V{s}}}\cdots \Ave{O_c}_{X_{\V{s}}}\ ,
\ee
where $X_{\V{s}}$ was defined in \Req{X-general}. Thus, 
\be
&&
\hat{S}-2\hat{T}+\hat{U}=\sum_{\V{s}}
\lbb 
\Ave{(\bar{p}_{\V{s}}^{a})^2(\bar{p}_{\V{s}}^{b})^2}_{\V{s}}
-2\Ave{(\bar{p}_{\V{s}}^{a})^2\bar{p}_{\V{s}}^{b}\bar{p}_{\V{s}}^{c}}_{\V{s}}
-\Ave{\bar{p}_{\V{s}}^{a}\bar{p}_{\V{s}}^{b}\bar{p}_{\V{s}}^{c}\bar{p}_{\V{s}}^{d}}_{\V{s}}
\rbb
\no \\
&&
=
\sum_{\V{s}}
\frac{1}{\int Dz X_{\V{s}}^n}\int Dz X_{\V{s}}^n
\lb \Ave{\bar{p}_{\V{s}}^2}_{X_{\V{s}}}-\Ave{\bar{p}_{\V{s}}}_{X_{\V{s}}}^2 \rb^2.
\ee
Noticing that
\be
\Ave{\bar{p}_{\V{s}}^2}_{X_{\V{s}}}-\Ave{\bar{p}_{\V{s}}}_{X_{\V{s}}}^2 =\Part{\log X_{\V{s}}}{\Lambda}{2}\ ,
\ee
the stability condition may be written, in the $n\to 0$ limit, under the form shown in \Req{ATcond-general}.

%%%%%%%%%%%%%%%%%%%%%%%%%%%%%%%%%%%%%%%%%%%%%%%%%%%%%%%%%%%%%%%%%%%%%%%%%%%%%%%%%%

\end{document}